\documentclass[twocolumn,numberedappendix, twocolappendix, apj]{openjournal}

\usepackage[english,english]{babel}
\usepackage{amsmath}
\usepackage[usenames]{color}
\usepackage{soul}
\usepackage{graphicx}
\usepackage{float}
\usepackage{bm}
\usepackage{comment}
\usepackage{xcolor}
\usepackage{enumitem}
\usepackage{physics}

\usepackage{longtable}
\usepackage{fontawesome5}

\usepackage{orcidlink}

\usepackage{hyperref}
\hypersetup{colorlinks=true, linkcolor=blue, citecolor=blue, filecolor=black, urlcolor=blue}

\begin{document}

\title{The Local Volume Database: a library of the observed properties of nearby dwarf galaxies and star clusters}

\author{Andrew B. Pace\,\orcidlink{0000-0002-6021-8760}$^{1,2,\star,\dagger}$ }
\thanks{$^\star$\href{mailto:pvpace1@gmail.com}{pvpace1@gmail.com} or \href{mailto:apace@virginia.edu}{apace@virginia.edu}}
\thanks{$^{\dagger}$ Galaxy Evolution and Cosmology (GECO) Fellow}
\thanks{
\faGithub \href{https://github.com/apace7/local_volume_database}{~Github}\newline 
\faBook \href{https://local-volume-database.readthedocs.io/}{~ReadTheDocs}\newline
\faChartLine \href{https://github.com/apace7/local_volume_database/tree/main/paper_examples}{~Paper~Plots}
}

\affiliation{$^{1}$ Department of Astronomy, University of Virginia, 530 McCormick Road, Charlottesville, VA 22904, USA}
\affiliation{$^{2}$ McWilliams Center for Cosmology, Carnegie Mellon University, 5000 Forbes Ave, Pittsburgh, PA 15213, USA }

\begin{abstract}
I present the Local Volume Database (LVDB), a catalog of the observed properties of  dwarf galaxies and star clusters in the Local Group and Local Volume. 
The LVDB includes positional, structural, kinematic, chemical, and dynamical parameters for  dwarf galaxies  and  star clusters.
I discuss the motivation, structure,  construction, and future expansion plans of the LVDB.
I highlight catalogs on faint and compact  ambiguous Milky Way systems, new Milky Way globular clusters and candidates, and globular clusters in nearby dwarf galaxies.
The LVDB is complete for known dwarf galaxies within $\sim 3 ~{\rm Mpc}$ and current efforts are underway to expand the database to resolved star systems in the Local Volume.
I present publicly available examples and use cases of the LVDB focused on the census and population-level properties of the Local Group and discuss some theoretical avenues.
The next decade will be an exciting era for near-field cosmology with many upcoming surveys and facilities, such as the Legacy Survey of Space and Time at the Vera C. Rubin Observatory, the Euclid mission, and the Nancy Grace Roman Space Telescope, that will both discover new dwarf galaxies and star clusters in the Local Volume and characterize known dwarf galaxies and star clusters in more detail than ever before.
The LVDB will be continually updated and is built to support and enable future dwarf galaxy and star cluster research in this data-rich era. 
The LVDB catalogs and package are publicly available as a GitHub repository, \texttt{local\_volume\_database}, and community use and contributions via GitHub are encouraged. 
\end{abstract}

 
\maketitle

\section{Introduction}
\label{sec:intro}

Dwarf galaxies are the most abundant type of galaxy in the universe.
The lowest mass dwarf galaxies are among the most chemically pristine and dark matter dominated galaxies known and they are relics from the early universe \citep[e.g.,][]{Tolstoy2009ARA&A..47..371T, Frebel2015ARA&A..53..631F, Simon2019ARA&A..57..375S, Battaglia2022NatAs...6..659B}. 
They are important probes of galaxy formation, reionization, and stellar feedback on the smallest scales \citep[e.g.,][]{Simon2019ARA&A..57..375S, Collins2022NatAs...6..647C}.
Only in the Local Group and Local Volume are the lowest mass dwarf galaxies observable and resolved into stars.

Over the past twenty years, the number of low mass dwarf galaxies and stellar systems in the Milky Way (MW), Local Group, and Local Volume has increased exponentially and pushed to fainter and fainter systems \citep[e.g.,][]{McConnachie2012AJ....144....4M, Muller2017A&A...597A...7M, Simon2019ARA&A..57..375S,  MutluPakdil2024ApJ...966..188M}.
Excitingly,  the next generation of upcoming wide-field photometric surveys, such as the Legacy Survey of Space and Time (LSST) at the Vera C. Rubin Observatory \citep{LSSTScienceCollaboration2009arXiv0912.0201L, Ivezic2019ApJ...873..111I}, the Euclid mission \citep{EuclidCollaboration2025A&A...697A...1E}, and the Nancy Grace Roman Space Telescope \citep{Spergel2015arXiv150303757S, WFIRST2019arXiv190205569A}, are on the cusp of starting and they are expected to discover a plethora of new dwarf galaxies \citep[e.g.,][]{Hargis2014ApJ...795L..13H, MutluPakdil2021ApJ...918...88M}.
Given the upcoming surveys, an updated census of both the Local Group and Local Volume will assist and enable research into near-field cosmology and nearby dwarf galaxies.
In particular, nearby dwarf galaxies are important cosmological probes to assess the many small-scale issues with our current $\Lambda$ cold dark matter ($\Lambda$CDM) model \citep[e.g.,][]{Bullock2017ARA&A..55..343B} and to uncover the nature of dark matter \citep[e.g.,][]{Buckley2018PhR...761....1B, Tulin2018PhR...730....1T, Strigari2018RPPh...81e6901S, Nadler2021PhRvL.126i1101N, McDaniel2024PhRvD.109f3024M}.

Almost forgotten, there are a number of faint stellar systems in the MW with sizes and luminosities that overlap with both the faint globular cluster population and the faint dwarf galaxy population without a clear classification into their nature \citep[e.g.,][]{Cerny2023ApJ...953....1C, Smith2024ApJ...961...92S}.
There is a renewed interest in these ambiguous systems as some may be the faintest and smallest galaxies \citep[e.g.,][]{Manwadkar2022MNRAS.516.3944M}, primordial star clusters \citep{Simon2024ApJ...976..256S}, and/or  their remnants \citep[e.g.,][]{Errani2024ApJ...968...89E}.
Classifying new satellites has been a difficult and time-consuming 
 problem in the MW, and classification will only get more difficult with the upcoming surveys.
As such,  an up-to-date census with the fundamental parameters of known stars clusters will be key to interpret new discoveries.

Here I introduce the Local Volume Database (LVDB), a catalog of nearby dwarf galaxies and star clusters resolved into individual stars. 
Section~\ref{section:lvdb} lays out the LVDB with subsections on the scope and purpose (\S~\ref{section:selection}), the content, structure, and construction (\S~\ref{section:content}), 
how to contribute (\S~\ref{section:contribution}), and citations (\S~\ref{section:citation}). 
Section~\ref{section:example} presents a series of examples and use cases of the LVDB highlighting the content of the LVDB and population level properties of nearby galaxies and star clusters. All examples are publicly available as \texttt{jupyter} notebooks on the GitHub repository.
Section~\ref{section:issues} discusses the current issues and limitations of the LVDB and plans for the expansion of the LVDB.
The LVDB is publicly available as a GitHub repository, \texttt{local\_volume\_database}\footnote{\url{https://github.com/apace7/local_volume_database}}. 

\section{The Local Volume Database}
\label{section:lvdb}

\subsection{Scope and Purpose}
\label{section:selection}

The primary goals of the LVDB are to: have a complete\footnote{I note that complete generally refers to whether the LVDB is complete for {\it known} systems.} catalog of {\it known} dwarf galaxies and star clusters in the Local Volume, have up-to-date fundamental observed properties of each system with errors and references, be able to easily add newly discovered systems and update any observed  property,  be able to reference the internal content of the catalog, and enable community contributions.
These goals are addressed in the LVDB structure. The LVDB is composed as a collection of \texttt{YAML}\footnote{YAML Ain't Markup Language} files, with the properties of an individual system in a single \texttt{YAML} file, which are combined together into user friendly catalogs. Second, the LVDB is hosted as a publicly available GitHub repository.

The LVDB aims to include high-quality measurements of the fundamental properties of both dwarf galaxies and star clusters.
While there are existing dwarf galaxy and star cluster catalogs and compilations no database includes both. One key reason to include both types of systems in the LVDB is the large number of existing small and faint systems that are difficult to classify between dwarf galaxies and stars clusters (i.e. ambiguous systems). Furthermore, these systems continue to be discovered in our current surveys. I expect the discovery rate to only increase with upcoming surveys (e.g., LSST).

Here, I refer to the Local Group, the Local Field\footnote{The edge of Local Group is generally defined at  the radius where the local Hubble flow starts and this  occurs at $R_{LG}\sim1~{\rm Mpc}$ \citep[e.g.,][]{Karachentsev2009MNRAS.393.1265K, Penarrubia2014MNRAS.443.2204P} and the Local Field is commonly defined as Local Group galaxies beyond the MW and M31. Given the catalogs in the LVDB, I define Local Field with an increased distance limit  ($d\lesssim 3~{\rm Mpc}$) as this corresponds to the distance limit that the LVDB is complete for {\it known} systems.}, and the Local Volume as systems  within $d\lesssim 1~{\rm Mpc}$,  within $d\lesssim 3~{\rm Mpc}$ but outside the influence of the MW or M31,  and  within $d\lesssim 10~{\rm Mpc}$, respectively. 
I follow \citet{McConnachie2012AJ....144....4M} in defining dwarf galaxies as systems fainter than $M_V \sim -18$. While brighter galaxies are included in the LVDB, they are primarily included to associate satellite dwarf galaxies with their host galaxy and to ensure the LVDB is complete in distance. The compiled properties of bright galaxies are limited, and other resources should be considered when studying the brighter galaxies ($M_V \lesssim -18$) in the Local Volume. Throughout this work, system will refer to a generic object within the LVDB (i.e., both dwarf galaxies and/or star clusters).

\subsection{Content and Structure} 
\label{section:content}

The LVDB is a compilation of dwarf galaxies and star clusters and their fundamental observed properties in the nearby universe. 
The starting point of the  dwarf galaxy  compilation is  the \citet{McConnachie2012AJ....144....4M} catalog\footnote{\url{https://www.cadc-ccda.hia-iha.nrc-cnrc.gc.ca/en/community/nearby/ }}. This has been supplemented by  two catalogs: the ``Catalog and Atlas of Local Volume galaxies'' \citep{Karachentsev2013AJ....145..101K}\footnote{\url{https://www.sao.ru/lv/lvgdb/}} and the ``Extragalactic Distance Database''
\citep{Tully2009AJ....138..323T, Anand2021AJ....162...80A}\footnote{\url{http://edd.ifa.hawaii.edu/ }} and subsequent discoveries.
At this point, the LVDB is complete   for known dwarf galaxies with $d<3~{\rm Mpc}$.
The star cluster compilation starts with the   \citet{Harris1996AJ....112.1487H} catalog and subsequent discoveries
and  is  supplemented by the more recent measurements compiled in the Galactic Globular Cluster Database\footnote{``Fundamental parameters of Galactic globular clusters''  \url{https://people.smp.uq.edu.au/HolgerBaumgardt/globular/} }  \citep{Baumgardt2018MNRAS.478.1520B, Baumgardt2020PASA...37...46B, Vasiliev2021MNRAS.505.5978V, Baumgardt2021MNRAS.505.5957B}. 
I have chosen not to include stellar streams in this catalog.  First, the \texttt{galstreams} package\footnote{\url{https://github.com/cmateu/galstreams}} \citep{Mateu2023MNRAS.520.5225M} contains an up-to-date list of MW stellar streams and their properties. Second, the properties of stellar streams are more difficult to characterize in the current LVDB structure. 

Previous   catalogs have used distance   \citep[$d \sim 3~{\rm Mpc}$;][]{McConnachie2012AJ....144....4M},  membership in the Local Volume  with a distance ($d<12 ~{\rm Mpc}$) and/or velocity selection ($v_{LG} < 600 {\rm km~s^{-1}}$) \citep{Karachentsev2013AJ....145..101K}, or  type (MW globular cluster or dwarf galaxy) to set their catalog selection. 
The LVDB selection is  to include dwarf galaxies and star clusters that are or can be resolved into individual stars. 
The effective limit of the LVDB is currently $d\lesssim10-12 ~{\rm Mpc}$, however, the dwarf galaxy component is only complete for known systems to $3~{\rm Mpc}$\footnote{The distance limit is continually updated, see GitHub release page documentation for the latest limit.} and the LVDB is complete for MW globular clusters and star clusters in nearby dwarf galaxies ($d\lesssim 1~{\rm Mpc}$). 
Given that space-based telescopes can resolve stars beyond the current LVDB distance limit ($d\sim 10-20~{\rm Mpc}$ with the Hubble Space Telescope and beyond $d\sim 40$ with James Webb Space Telescope imaging), I anticipate that the LVDB distance limit will increase in the future.

The LVDB is structured as a collection of \texttt{YAML} files  with each system in its own \texttt{YAML} file. The \texttt{YAML} files are then combined into catalogs with additional value-added columns.  
The measurements compiled in the \texttt{YAML} files include: positions, classification, structural parameters, distance, luminosity, stellar kinematics, stellar chemistry, systemic proper motion, HI gas content, and mean age. 
The properties and measurements in the input \texttt{YAML} files and the value-added columns are described in detail in Appendix~\ref{appendix:yaml}.

\subsubsection{Primary Catalogs}

The primary catalogs of the LVDB are split by host and system classification. The  primary catalogs are as follows:

\begin{itemize}
    \item MW dwarf galaxies.
    \item M31 dwarf galaxies.
    \item Local Field dwarf galaxies ($d \lesssim 3~{\rm Mpc}$) unassociated with the MW or M31 \citep[i.e., an updated][compilation]{McConnachie2012AJ....144....4M}.
    \item Local Volume dwarf galaxies ($ d\gtrsim 3$ Mpc).
    \item Ambiguous systems in the MW (generally in the Galactic halo; $\abs{b}\gtrsim10^{\circ}$; labeled \texttt{gc\_ambiguous}). 
    Referred to as ambiguous or hyper-faint\footnote{I use hyper-faint to refer to $M_V > -3$ \citep{Hargis2014ApJ...795L..13H}, although Hargis et al. use $L_V<10^3 ~L_{\odot}$ or $M_V \sim -2.7$ to refer to hyper-faint.} compact stellar systems here (HFCSS; see Section~\ref{section:ufcss} for a more thorough discussion).
    \item Globular clusters (GCs) in the \citet{Harris1996AJ....112.1487H} catalog.
    \item Recently discovered globular clusters   (i.e. post-Harris catalog; labeled \texttt{gc\_mw\_new}). These are primarily located at low Galactic latitude ($\abs{b}\lesssim10-20^{\circ}$) but a few systems are located at high Galactic latitudes. Referred to as GC new or GC new bulge/disk/halo  here. 
    \item Globular clusters and star clusters in nearby dwarf galaxies (labeled \texttt{gc\_dwarf\_hosted}). This includes the LMC/SMC globular clusters and star clusters but not any of the  globular clusters associated with the Sagittarius dwarf galaxy as they are included in the \citet{Harris1996AJ....112.1487H} catalog. Referred to as GC dwarf hosted here.
    \item Lower-quality candidates, confirmed false positive systems, and background systems. The background  systems compiled here were initially considered satellites or members of the  Local Volume. 
\end{itemize}

\noindent The  dwarf galaxy catalogs with $d \lesssim 3~{\rm Mpc}$ are complete for {\it known} dwarf galaxies. 
The  Local Volume compilation is not complete for known systems and currently focuses on  dwarf galaxies that either host globular clusters  or are members of one of the closest groups (e.g., Cen A (NGC 5128), NGC 253 (Sculptor), and M 81 (NGC 3031)).

The separation of globular cluster catalogs is based on classification, host, and the common interloper type. 
The ambiguous systems have a distinct catalog due to their unclear classification between as star cluster or dwarf galaxy. 
The new Galactic disk/bulge/halo star clusters are split from the Harris catalog as candidates at low Galactic latitude require follow-up spectroscopy to be confirmed due to the higher stellar foreground and open clusters represent an interloper population at low Galactic latitudes.
The new Galactic disk/bulge/halo catalog  may not be  complete for new literature candidates.
The dwarf galaxy globular cluster table is not complete and there are many globular cluster candidates  that require (spectroscopic) confirmation. 
Lastly, the candidate/false positive table is not complete and is partly included for any current system in the LVDB that may be classified as a false positive in the future.

\subsubsection{Classification}
\label{sec:classification}

The LVDB contains two general types of systems (dwarf galaxy and star cluster) and classification is key component of the LVDB. 
The inclusion of a system in a table has an implicit  classification and many systems  are not yet confidentially classified (generally some combination of faint, small, and/or distant). 
I follow  \citet{Willman2012AJ....144...76W} to define a galaxy: ``A galaxy is a gravitationally bound collection of stars whose properties cannot be explained by a combination of baryons and Newton’s laws of gravity.'' 
In the context of $\Lambda$CDM, the galaxy definition implies that the system has  a dark matter halo.
I use the following to define a star cluster: ``A star cluster is a  self-gravitating bound collection of stars.''
There are  several classification columns in the LVDB that summarize these definitions.

The following  methods are used to classify systems as dwarf galaxies or star clusters: 
\begin{itemize}
    \item dynamical mass-to-light ratio---The  gold standard   and most direct method  to infer the presence (or absence) of a dark matter halo is to measure the  dynamical mass-to-light ratio. For a system with a dark matter halo, the dynamical mass-to-light ratio will be larger than the prediction for a pure stellar system ($\Upsilon_{\rm dyn} > {\rm few} \times \Upsilon_{\star}$). 
    For pressure-supported systems, such as a MW dwarf spheroidal galaxy, this is commonly done by measuring the stellar velocity dispersion from radial velocity of individual stars. 
    For faint systems, the predictions for pure stellar systems are  small ( $\sigma_{\star}\lesssim 0.1~{\rm km~s^{-1}}$) and a dark matter halo is inferred if the  velocity dispersion is resolved and non-zero.  For low mass systems ($\sigma_v \lesssim 5~{\rm km~s^{-1}}$) with small sample sizes ($N\lesssim 20$), the velocity dispersion can be inflated by unresolved binary stars  \citep{McConnachie2010ApJ...722L.209M, Minor2010ApJ...721.1142M, Minor2019MNRAS.487.2961M}. 
    \item stellar metallicity dispersion---An observed stellar metallicity spread implies the system was able to retain supernova  ejecta to self-enrich the next generation of star formation. This indirectly infers the system has or had a dark matter halo to retain the enriched gas \citep[][and references therein]{Willman2012AJ....144...76W}.
    \item size---Star clusters have smaller sizes ($R_{1/2}\lesssim 10-20~{\rm pc}$) than dwarf galaxies and a large size ($R_{1/2}\gtrsim 50~{\rm pc}$) would infer a dwarf galaxy classification. I note that two recently discovered  globular clusters, Laevens~1/Crater~I and Sagittarius~II, have relatively large sizes   \citep[$R_{1/2}\sim20-30~{\rm pc}$;][]{Weisz2016ApJ...822...32W, Longeard2021MNRAS.503.2754L, Richstein2024ApJ...967...72R}. However, this method stops working for faint luminosities ($M_V\gtrsim-3$), as the star cluster and dwarf galaxy populations overlap in their size distributions.
    \item mass segregation---The presence of mass segregation in a system can classify a system as a star cluster  \citep[e.g.,][]{Baumgardt2022MNRAS.510.3531B}. High mass stars are expected to sink to the centers of the systems and low mass stars pushed to the outskirts. The presence of a dark matter halo significantly increases the timescales for mass segregation  \citep{Baumgardt2022MNRAS.510.3531B}. Old stellar systems with low relaxation times  should be mass segregated \citep[e.g.,][]{Forbes2011PASA...28...77F, Weatherford2018ApJ...864...13W, Weatherford2020ApJ...898..162W, Baumgardt2022MNRAS.510.3531B}.   
    Systems with a relaxation time longer than the age of the universe can be classified as galaxies whereas systems with shorter times would be star clusters \citep{Forbes2011PASA...28...77F}. Observationally, mass segregation has been quantified by measuring $R_{1/2}$ of bright and faint stars and comparing the measurement to N-body simulations tailored to the system's properties \citep[e.g.,][]{Baumgardt2022MNRAS.510.3531B}.  
    \item light element abundances---Globular clusters are observed to have light element variations. These variations are  generally either correlated (e.g., Na-Ne) or anti-correlated  (e.g., Na-O and Mg-Al)  \citep[e.g.,][]{Carretta2009A&A...505..139C}.  Identifying a system with  correlated (or anti-correlated) light element variations can be used to classify a system as a globular cluster.  However, the enriched fraction has been found to depend on the initial stellar mass of the globular cluster \citep[e.g.,][]{Gratton2019A&ARv..27....8G} and many systems are below this threshold (including all of the ambiguous systems).  
    \item neutron capture elements---Ultra-faint dwarf galaxies and globular clusters are observed to have different  abundances of   neutron-capture elements (e.g., Sr, Ba, and Eu)  \citep{Ji2019ApJ...870...83J}. In general, ultra-faint dwarf galaxies are deficient in  neutron-capture elements \citep{Ji2019ApJ...870...83J} whereas globular clusters follow the MW  halo trend. The low neutron-capture elements in ultra-faint dwarf galaxies may be due to stochastic enrichment, metal loss in winds, and short star formation durations \citep{Ji2019ApJ...870...83J}. There are several exceptions to the ultra-faint dwarf galaxy trend which may be evidence for rare  enrichment events: Reticulum~II \citep{Ji2016Natur.531..610J, Roederer2016AJ....151...82R}, Tucana~III \citep{Hansen2017ApJ...838...44H, Marshall2019ApJ...882..177M}, and Grus~II \citep{Hansen2020ApJ...897..183H}. 
\end{itemize}

Lastly, I note that there are some additional  types of systems with some ambiguity in their classification: nuclear star clusters, extended clusters, ultra-compact dwarf (galaxies), and tidal dwarf (galaxies). The two former systems are classified here as star cluster here but may contain metallicity spreads or have large sizes and the two latter  systems are classified as dwarf galaxies here but may have small sizes or low dynamical masses. See \citet{Forbes2011PASA...28...77F, Willman2012AJ....144...76W} for more discussion on the classification of these other types of systems.

\subsection{Contributions}
\label{section:contribution}

Community contributions to the LVDB are welcomed and encouraged. 
While the current content of the LVDB has been collected and curated by myself,  community contributions can be easily included by hosting the LVDB as a  GitHub repository.
Some   common anticipated contributions may be    adding new systems or adding new measurements of existing systems.
In both cases this would consist of adding or editing input \texttt{YAML} files and submitting a GitHub pull request to merge the content into the main branch\footnote{This would include validation and confirmation for pull requests by myself or other LVDB GitHub maintainers.}. 
The detailed description of the input \texttt{YAML} files is in Section~\ref{appendix:yaml} and due to the structure of the LVDB, new types of measurements  can be added as new \texttt{YAML} keys. 
Section~\ref{section:issues} discusses some current issues and limitations along with a number expansion ideas that are excellent starting areas for anyone interested in contributing.

\begin{figure}
\includegraphics[width=\columnwidth]{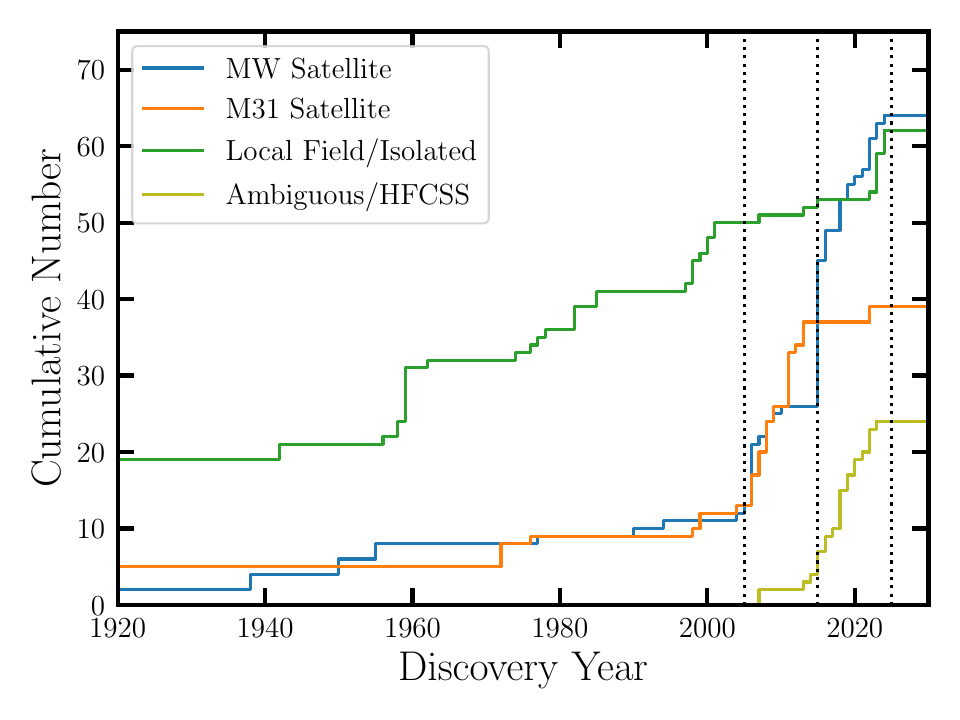}
\caption{The census of dwarf galaxies in the Local Field as a function of discovery year. The cumulative distributions show the  number of MW dwarf galaxies  (blue), M31 dwarf galaxies (orange),  Local Field (green), and ambiguous or hyper-faint compact stellar systems (olive).  The three dotted lines  at 2005, 2015, 2025  represent the three eras of digital sky surveys:  the past SDSS era;  the ongoing era with Blanco/DECam, Pan-STARRS, ATLAS, {\it Gaia}, Subaru/HSC, CFHT/Megacam;  the upcoming era with  Vera C. Rubin/LSST, Euclid, and Nancy Grace Roman Space Telescope. 
}
\label{fig:discovery_year}
\end{figure}

\begin{figure}
\includegraphics[width=\columnwidth]{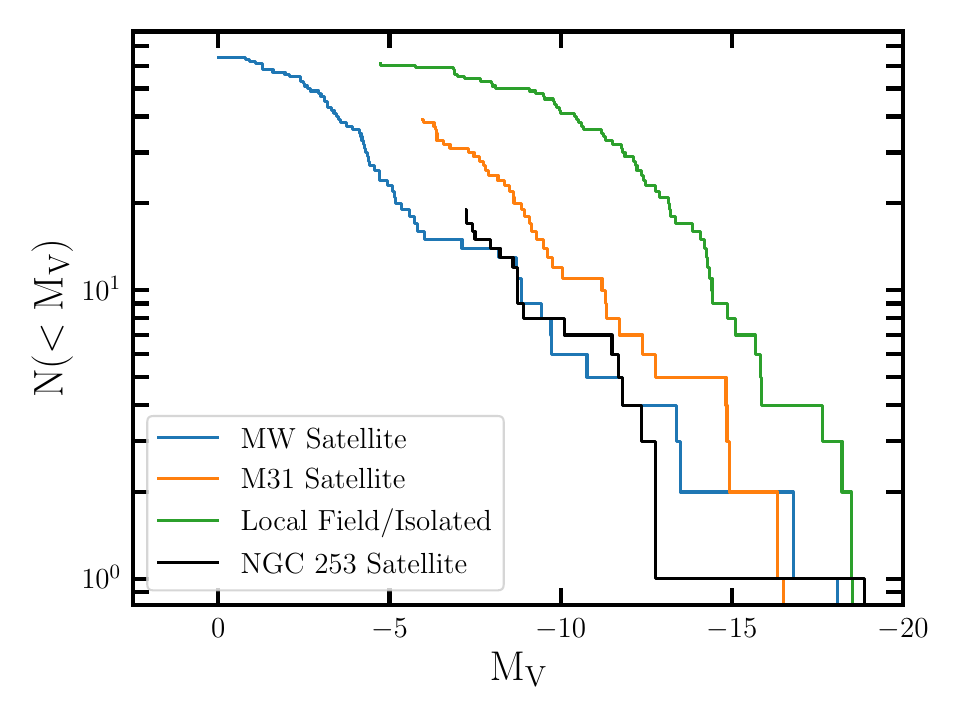}
\caption{ The observed cumulative dwarf galaxy luminosity function of the MW (blue), M31 (orange), the Local Field (green), and NGC~253 (black). 
}
\label{fig:cumulative_distribution}
\end{figure}

\begin{figure*}
\begin{center}
\includegraphics[width=\textwidth]{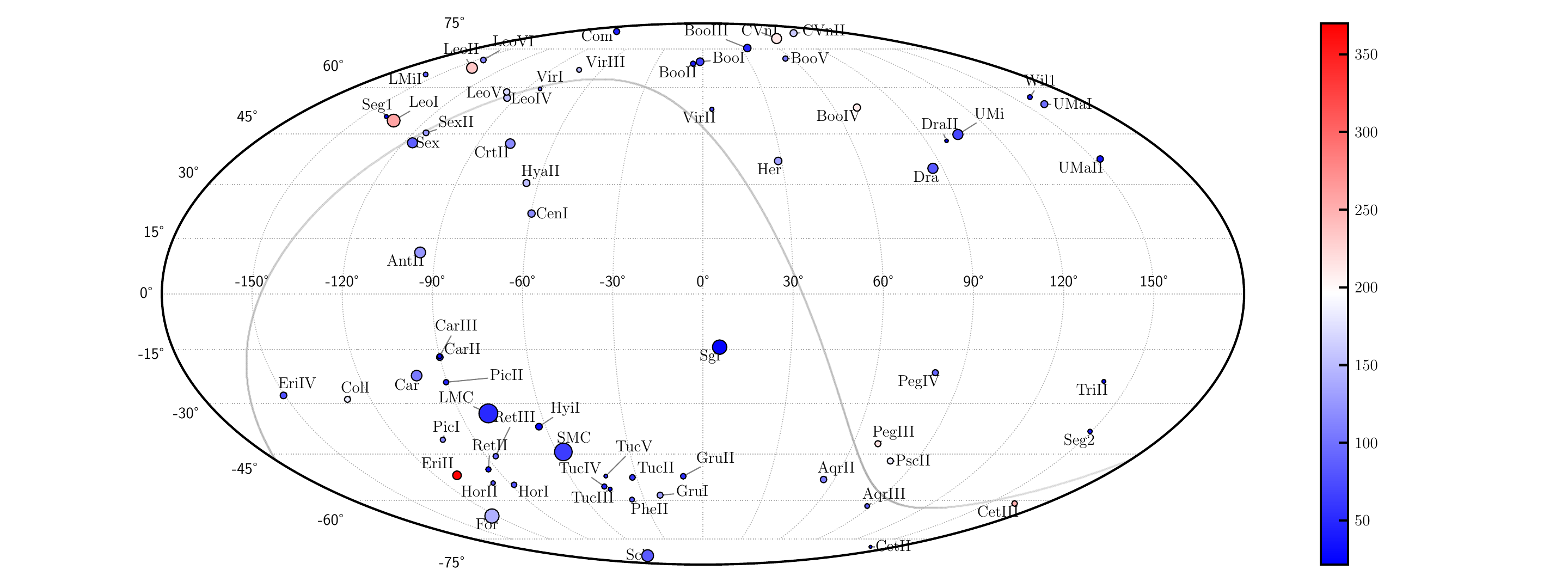}
\includegraphics[width=0.75\textwidth]{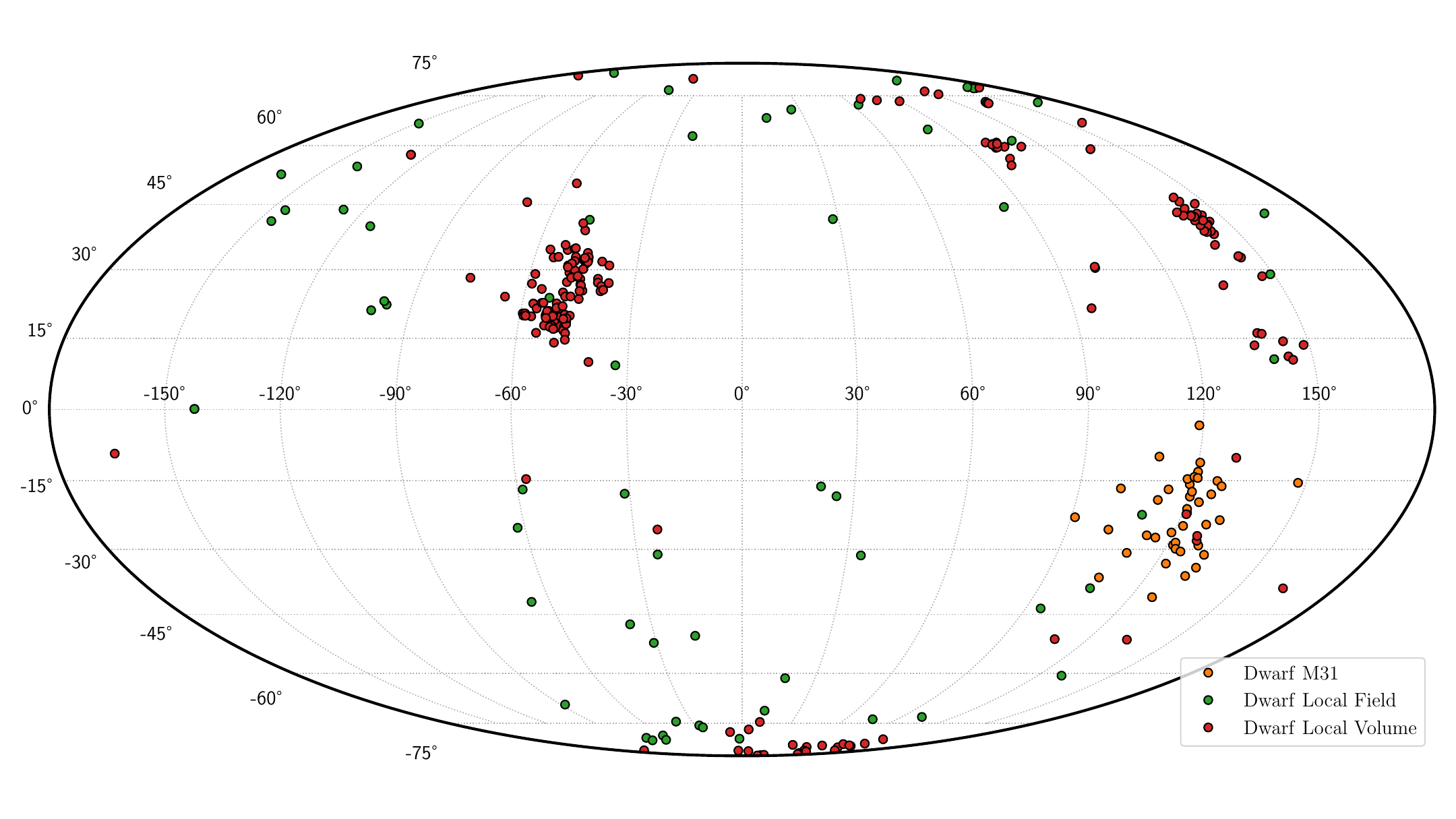}
\includegraphics[width=0.75\textwidth]{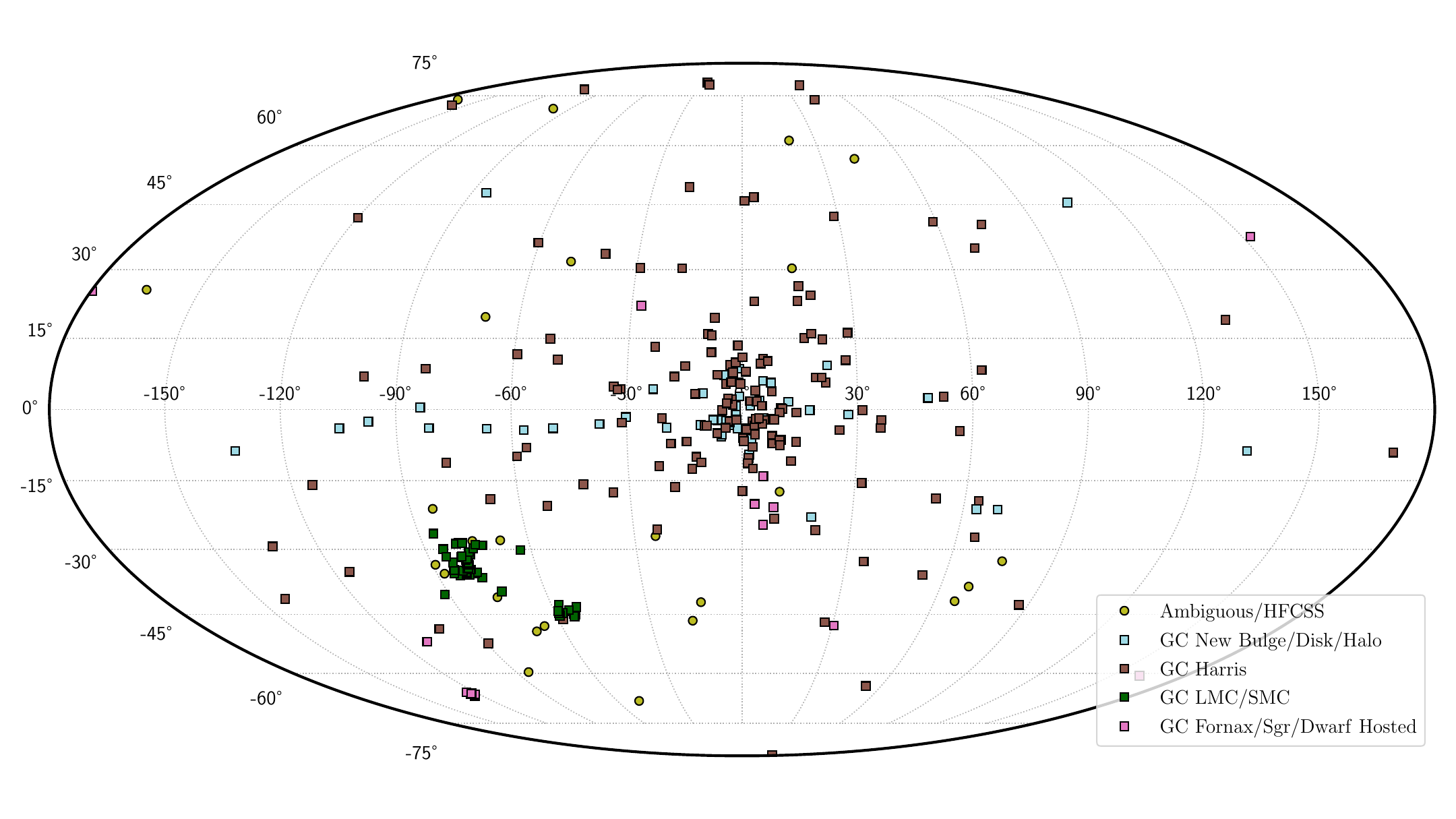}
\caption{Aitoff projection in Galactic coordinates of dwarf galaxies and globular clusters in the LVDB. {\bf Top}: The MW dwarf galaxy population. The size of the points is proportional to the system's stellar mass and the colorbar shows the system's heliocentric distance. Each system is labeled by its abbreviation.  For reference, a line is included at $\delta=0^{\circ}$. {\bf Middle}: Dwarf galaxies outside the MW including: M31 (orange),  Local Field (green), and  Local Volume (red).  {\bf Bottom}: Milky Way star clusters: ambiguous or HFCSS (olive), globular clusters in the \citet{Harris1996AJ....112.1487H} catalog (brown), newly discovered globular clusters or candidates (light blue), LMC/SMC star clusters (dark green), and star clusters in dwarf galaxies including star clusters associated with the Fornax and Sagittarius dwarf galaxies (pink).  
}
\label{fig:aitoff_projection}
\end{center}
\end{figure*}

\subsection{Citations}
\label{section:citation}

As the LVDB relies on a significant amount of input literature, there are several tools that enable citations to the  internal content of the LVDB. 
Users are encouraged to cite both the LVDB and the internal content used in their analysis.
All measurements have the corresponding reference included.  The reference columns follow a consistent schema of author last name + Astrophysics Data System (ADS) bibcode\footnote{https://ui.adsabs.harvard.edu/help/actions/bibcode}.  The author's last name is stripped of special characters and spaces but capitalization is kept. While the ADS bibcode is a unique identifier for a reference, the addition of the author's last name significantly improves the human interpretability of the LVDB reference schema.
To aid in enabling references to the internal contents of the database, there is an up-to-date \texttt{BibTeX} file containing all entries in the LVDB with the LVDB reference schema\footnote{ \url{https://github.com/apace7/local_volume_database/blob/main/table/lvdb.bib}} and there is an ADS library with LVDB references\footnote{\url{https://ui.adsabs.harvard.edu/public-libraries/fVKkEJbdRyCmscCOwzsz6w}}.
As the ADS bibcode has a fixed length of 19 characters, the ADS bibcode can be derived from the last 19 characters of the LVDB reference schema.
There are several   \texttt{jupyter} notebooks that show examples and assist with  LVDB internal citations. 

Lastly, users should consider acknowledging the other catalogs and databases that the LVDB was built off of. For dwarf galaxies, this includes the \citet{McConnachie2012AJ....144....4M} catalog, the ``Catalog and Atlas of Local Volume galaxies'' \citep{Karachentsev2013AJ....145..101K}, and the ``Extragalactic Distance Database'' 
\citep{Tully2009AJ....138..323T, Anand2021AJ....162...80A}.
For globular clusters this includes: the \citet{Harris1996AJ....112.1487H} catalog, and the Galactic Globular Cluster Database \citep{Baumgardt2018MNRAS.478.1520B, Baumgardt2020PASA...37...46B, Vasiliev2021MNRAS.505.5978V, Baumgardt2021MNRAS.505.5957B}.

\section{Use Cases: The Properties of Dwarf Galaxies and Star Clusters in the Local Group and Beyond}
\label{section:example}

To show the content  of the LVDB and what it can be used for, I present a series of use cases and examples. \texttt{jupyter} notebooks  that produce the all of the figures are publicly available in the LVDB GitHub repository\footnote{\url{https://github.com/apace7/local_volume_database/tree/main/paper_examples/}}.

\subsection{Discovery and Census}
The number of dwarf galaxies in the Local Volume has significantly expanded since the start of large digital sky surveys. 
Figure~\ref{fig:discovery_year} shows the number of known or candidate dwarf galaxies as a function of time for the MW, M31, Local Field, and the ambiguous/HFCSS MW systems. 
For MW dwarf galaxies, each generation of sky surveys has roughly doubled the number of known satellites, and the next generation of surveys is on the cusp of starting.
The initial set of MW dwarf galaxies were discovered on photographic plates\footnote{This excludes the LMC and SMC which can be seen by eye.} (these systems are commonly referred to as the `classical' dwarf spheroidal galaxies) and only the highest surface brightness dwarf galaxies were discovered \citep[e.g.,][]{Shapley1938Natur.142..715S, Harrington1950PASP...62..118H, Irwin1990MNRAS.244P..16I}. 
With the advert of digital CCDs and the start of the Sloan Digital Sky Survey (SDSS), the number of known  dwarf galaxies roughly doubled (from 11 to 26) and the discovery space was pushed into the ultra-faint regime\footnote{Defined as $M_V > -7.7$ following \citet{Simon2019ARA&A..57..375S}.} \citep[e.g.,][]{Willman2005AJ....129.2692W, Walsh2007ApJ...662L..83W, Belokurov2007ApJ...654..897B, Belokurov2010ApJ...712L.103B}. 
The next generation of wide field surveys (what I refer to as the current era here) has increased the number of dwarf galaxies to at least $65$. 
This era has been driven  by several new instruments on existing telescopes including: DECam on the Blanco telescope, the  Panoramic Survey Telescope and Rapid Response System (Pan-STARRS), and the Hyper Suprime-Cam on the  Subaru telescope. DECam has been utilized in multiple  surveys that have been used to discover new MW satellites such as  the Dark Energy Survey \citep[DES; e.g.,][]{Bechtol2015ApJ...807...50B, DrlicaWagner2015ApJ...813..109D}, the DECam Local Volume Exploration Survey  \citep[DELVE; e.g.,][]{Mau2020ApJ...890..136M, Cerny2023ApJ...953....1C}, and several smaller programs \citep[e.g.,][]{Kim2015ApJ...803...63K, Torrealba2018MNRAS.475.5085T}.
The initial Pan-STARRS survey found several new systems \citep[e.g.,][]{Laevens2014ApJ...786L...3L, Laevens2015ApJ...813...44L}, and is now part of the  Ultraviolet Near Infrared Optical Northern Survey \citep[UNIONS;][]{Smith2023AJ....166...76S, Smith2024ApJ...961...92S}.
Lastly, the Hyper Suprime-Cam Subaru Strategic Survey was used  to discover several new MW satellites \citep[e.g.,][]{Homma2019PASJ...71...94H, Homma2024PASJ...76..733H} and its data quality is a preview of the data quality in the upcoming LSST.

The Milky Way dwarf galaxy census is now partly limited by the ability to classify faint satellites and there are over 30 satellites without a secure classification.
The number of the ambiguous systems or HFCSS  has  grown in recent years due to sensitive wide-field surveys. Due to the structure of the LVDB, the current ambiguous systems will move to different catalogs when they are confidently classified.

Similarly, the initial M31 satellite dwarf galaxies  were discovered on photographic plates \citep[e.g.,][]{vandenBergh1972ApJ...171L..31V, Armandroff1998AJ....116.2287A, Karachentsev1999A&A...341..355K}. 
The M31 dwarf galaxy  population has since benefited from all sky  surveys such as:
SDSS \citep[e.g.,][]{Zucker2004ApJ...612L.121Z, Bell2011ApJ...742L..15B}, Pan-STARRS \citep{Martin2013ApJ...772...15M, Martin2013ApJ...779L..10M}, and The Dark Energy Camera Legacy Survey \citep[DECaLS;][]{Collins2022MNRAS.515L..72C, MartinezDelgado2022MNRAS.509...16M}.
Dedicated deep, wide field surveys such  as the Pan-Andromeda Archaeological Survey (PandAS) with CFHT/Megacam has discovered a number of M31 satellites  \citep[e.g.,][]{Martin2006MNRAS.371.1983M, Ibata2007ApJ...671.1591I, Richardson2011ApJ...732...76R}. 
Many more satellite dwarf galaxies are predicted to exist in the M31 system \citep{DolivaDolinsky2023ApJ...952...72D} but M31 is outside the LSST and Euclid footprints.

While the current surveys have yet to build a complete census of  dwarf galaxies within their reach, the next generation of wide-field surveys is about to begin. 
The first data release of the 10-year LSST at the Vera C. Rubin Observatory is planned for 2026. 
There are multiple  projections for the total MW dwarf galaxy population to be  $N\gtrsim100-200$ \citep[e.g.,][]{Hargis2014ApJ...795L..13H, Newton2018MNRAS.479.2853N, Manwadkar2022MNRAS.516.3944M, Nadler2024ApJ...967...61N, Ahvazi2024MNRAS.529.3387A} and  LSST  is expected significantly increase number of known MW dwarf galaxies in its footprint.
The Euclid mission launched on July 1, 2023 and  the first data release is scheduled for  2026 and the Nancy Grace Roman Space Telescope is planned to launch between Oct 2026 to May 2027. While both space telescopes can assist in the MW dwarf galaxy census, their new and unique discovery regime with resolved stars will be low mass dwarf galaxies in the Local Volume  \citep[for example, Euclid has already discovered a dwarf galaxy around the MW-analog, NGC 6744 at $d\sim9.5~{\rm Mpc}$;][]{Hunt2025A&A...697A...9H}.

Figure~\ref{fig:cumulative_distribution} shows the observed cumulative luminosity function of dwarf galaxies in the MW, M31,   Local Field, and NGC~253\footnote{Of the MW-like galaxies beyond $d>3~{\rm Mpc}$,  NGC~253 is the only host in the LVDB that is complete for its known satellite galaxy population.} systems. 
Beyond the MW,  dwarf galaxy discoveries are just hitting  the ultra-faint  regime   and the next generation of surveys and telescopes (e.g., Rubin/LSST, Euclid, Roman Space Telescope) will have the sensitivity to probe this dwarf galaxy regime. 
Already, the LVDB is starting to add the dwarf galaxy populations of MW-like systems in the Local Volume (e.g., Cen~A, NGC 253, M81, M101). 

The dwarf galaxy luminosity function is a key observable in near-field cosmology that probes both the nature of dark matter and galaxy formation physics at low halo masses.
Initial research focused on the total number of satellites and whether the number of MW satellite dwarf galaxies was consistent with $\Lambda$CDM predictions \citep[e.g.,][]{Klypin1999ApJ...522...82K, Moore1999ApJ...524L..19M, Koposov2008ApJ...686..279K, Tollerud2008ApJ...688..277T}.
More recent studies have shown that the total number is consistent with $\Lambda$CDM predictions \citep{Kim2018PhRvL.121u1302K},  used the dwarf galaxy population to study the minimum halo mass that can host galaxy \citep{Nadler2020ApJ...893...48N}, and placed constraints on different dark matter models \citep{Nadler2021PhRvL.126i1101N}. 
Similarly, the  satellite luminosity function of other MW-like hosts such as  M31 \citep{DolivaDolinsky2023ApJ...952...72D} and
Centaurus~A \citep{Weerasooriya2024ApJ...968...78W}   can be used to study galaxy  formation physics and test environmental differences for MW-like hosts throughout the Local Volume.

\begin{figure*}
\includegraphics[width=0.95\textwidth]{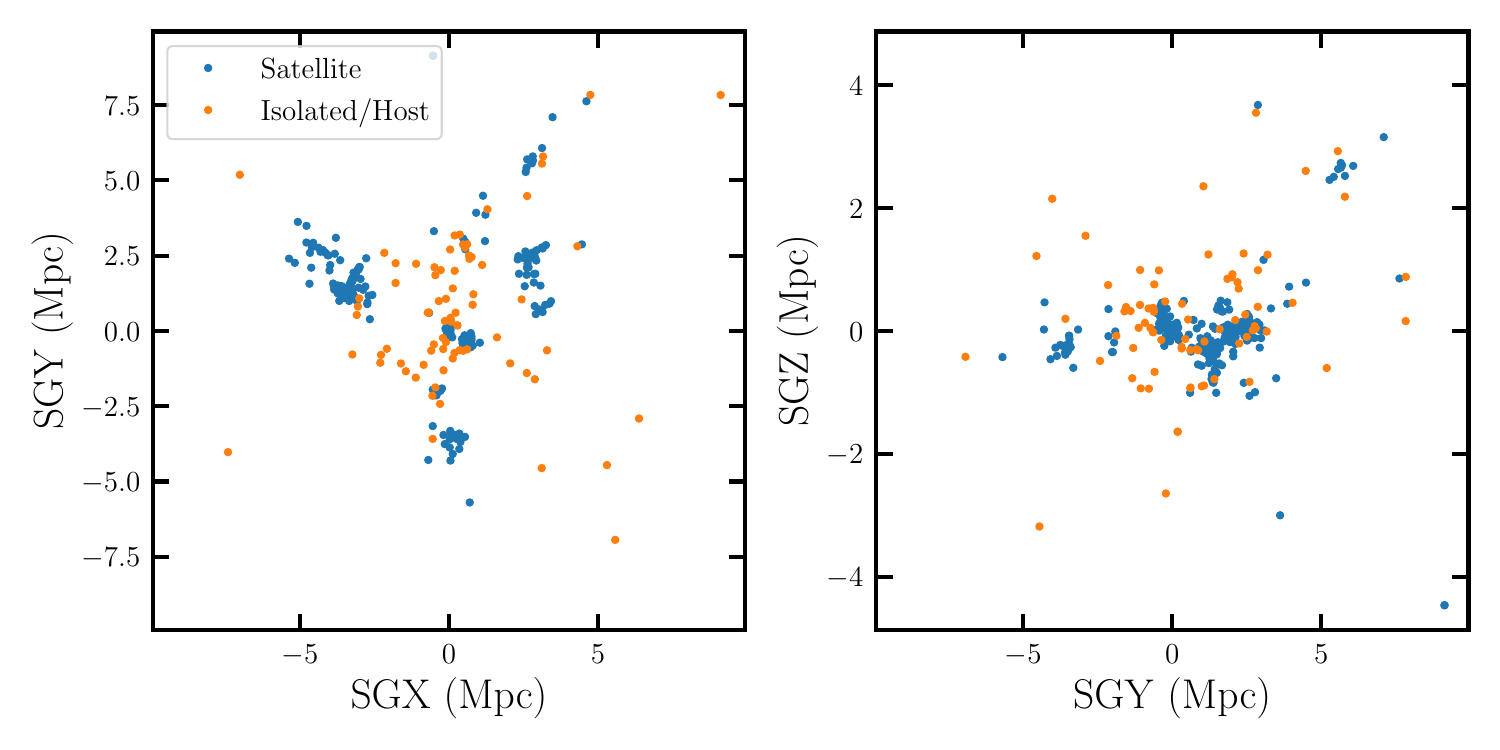}
\caption{The dwarf galaxies in the LVDB in Supergalactic coordinates. The dwarf galaxies are separated by whether they are a satellite (blue) or isolated/host (orange). 
}
\label{fig:supergalactic}
\end{figure*}

\subsection{Spatial Distribution}

The top panel of Figure~\ref{fig:aitoff_projection} shows Aitoff projections in Galactic coordinates of the MW dwarf galaxy satellites. 
There are a number of spatial alignments and combined with the systemic motion (velocity and proper motion), the phase-space properties of the satellite system can be studied \citep[e.g.,][]{Pawlowski2021Galax...9...66P}.
There is a clear overdensity of dwarf galaxies around LMC and SMC in Figure~\ref{fig:aitoff_projection}. This is the closest example of group infall and a number of  satellites have been identified that are or were associated with the LMC and/or SMC   \citep[e.g.,][]{Kallivayalil2018ApJ...867...19K, Erkal2020MNRAS.498.5574E, Patel2020ApJ...893..121P, CorreaMagnus2022MNRAS.511.2610C, Pace2022ApJ...940..136P}.
Other examples of  spatial alignments are the candidate  Crater-Leo group \citep[Leo~II, Leo~IV, Leo~V, and Crater~I;][]{Torrealba2016MNRAS.459.2370T, Fritz2018A&A...619A.103F, Julio2024A&A...687A.212J}, the vast polar structure \citep[e.g.,][]{LyndenBell1976MNRAS.174..695L, Pawlowski2020MNRAS.491.3042P, Pawlowski2021Galax...9...66P}, and satellite pairs (e.g., Draco-Ursa~Minor, Pisces~II-Pegasus~III, Carina~II-Carina~III, Leo~IV-Leo~V).
All the spatial structure in the MW satellite population has the caveat that there are  large selection effects along the Galactic Plane.

The central panel of Figure~\ref{fig:aitoff_projection} shows  Aitoff projections in Galactic coordinates of all the  dwarf galaxies in the LVDB beyond the MW. There are several overdensities that correspond to the satellite systems of M31, NGC 253 (Sculptor),  Centaurus A (NGC 5128), and M 81 (NGC 3031).  
Many new dwarf galaxies and candidates have been recently uncovered around NGC 253 \citep{Sand2014ApJ...793L...7S, MartinezDelgado2024arXiv240503769M} and  Centaurus A \citep{Crnojevic2014ApJ...795L..35C, Muller2017A&A...597A...7M} due to   dedicated deep imaging campaigns and all sky surveys (e.g., DECaLS). 
The Local Field and Local Volume are areas where upcoming surveys such as Rubin/LSST, Euclid, and Roman will enable the discovery of faint and distant dwarf galaxies  \citep[e.g.,][]{MutluPakdil2021ApJ...918...88M}.

The bottom panel of Figure~\ref{fig:aitoff_projection} shows the globular and star cluster population of the MW including globular clusters hosted by  MW dwarf galaxy satellites (e.g.  LMC,  SMC, Fornax, Sagittarius). There are many new  confirmed and candidate  globular  clusters at low  Galactic latitudes that have been discovered with  near-infrared surveys \citep[e.g.,][]{Minniti2011A&A...527A..81M, Garro2020A&A...642L..19G} and/or with {\it Gaia} astrometry \citep[e.g.,][]{Gran2022MNRAS.509.4962G, Pace2023MNRAS.526.1075P}. 
Over 30 new confirmed or candidate globular clusters  have been uncovered in the Galactic bulge in recent years \citep{Bica2024A&A...687A.201B, Garro2024A&A...687A.214G}.
Since the start of wide-field imaging surveys targeting high Galactic latitude, there are a number of new compact stellar systems  in the Galactic halo and while some are clearly star clusters \citep[e.g., Laevens~1/Crater~I and Sagittarius~II;][]{Weisz2016ApJ...822...32W, Longeard2021MNRAS.503.2754L} the majority are not confidentially classified  \citep[e.g.,][]{Koposov2007ApJ...669..337K, Cerny2023ApJ...953....1C}.
Several of the HFCSS are likely associated with the LMC and/or SMC \citep[e.g.,][]{Martin2016ApJ...830L..10M, Gatto2021RNAAS...5..159G, Cerny2023ApJ...953L..21C}.
I note that the LVDB  is only complete for  old LMC and SMC star clusters. 
While the population of bright globular clusters at high Galactic latitude ($\lvert b\rvert >10^{\circ}$) is generally thought to be complete, the orbital configuration of the current globular clusters suggest there may be 1-3 missing globular clusters in the outer halo \citep{Webb2021MNRAS.502.4547W}.

Figure~\ref{fig:supergalactic} shows the Local Volume dwarf galaxies  in supergalactic coordinates. The systems are  separated by whether they are a satellite or isolated. 
As the LVDB is not yet complete for  distant Local Volume galaxies, the Supergalactic plane does not stand out.

\begin{figure*}
\includegraphics[width=\textwidth]{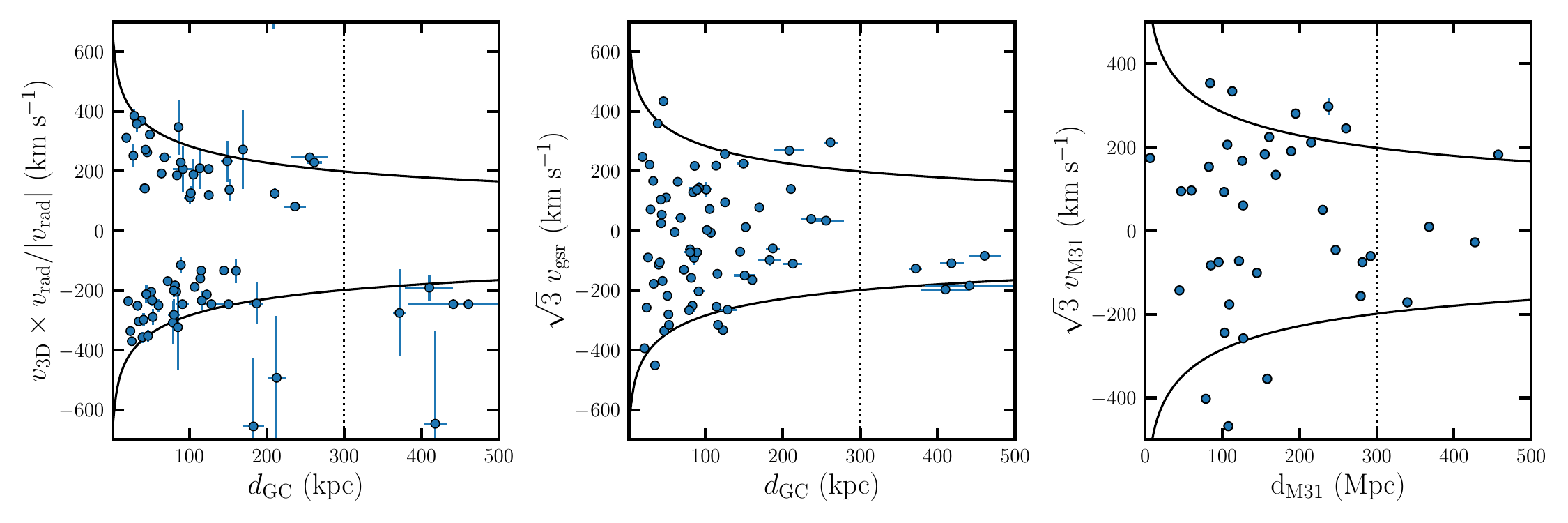}
\caption{Distance versus velocity for the dwarf galaxies in the MW and M31 systems. {\bf Left}: Distance to the Galactic center versus total 3D velocity, where positive (negative) indicates whether the system is moving away from (towards) the Galactic center  for MW dwarf galaxies. The solid lines denote the escape velocity curve of the MW with the \texttt{MWPotential2014} from \texttt{galpy}. The dotted line denotes the approximate virial radius at $300~{\rm kpc}$. {\bf Middle}: Distance to the Galactic center versus corrected Galactocentric velocity for MW dwarf galaxies. A factor of $\sqrt{3}$ is included to account for the tangential velocity and show the approximate total velocity prior to {\it Gaia} astrometry. 
{\bf Right}: Velocity and distance relative to M31 for the M31 dwarf galaxies. A similar factor of $\sqrt{3}$ is included  to account for the unknown tangential velocity.  The same escape velocity and virial radius are assumed for M31 even though more recent mass measurements suggest M31 is more massive than the MW. 
}
\label{fig:radius_velocity}
\end{figure*}

\begin{figure}
\includegraphics[width=\columnwidth]{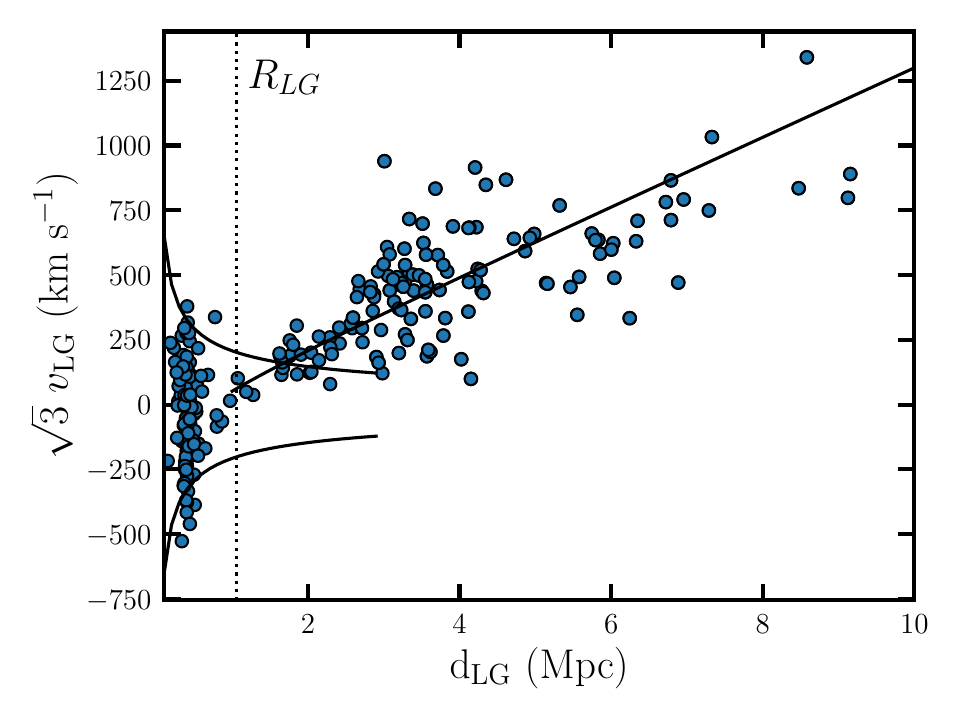}
\caption{Distance versus (corrected) velocity relative to the Local Group for  dwarf galaxies in the Local Volume.   The solid curves show the escape velocity curves of a point mass with $M_{\rm LG}=5\times10^{12}~{\rm M_{\odot}}$ and are cut off at 3 Mpc.  The dotted line is the Local Group turnaround radius, $R_{LG}\sim1~{\rm Mpc}$ \citep{Karachentsev2009MNRAS.393.1265K}. The solid line shows the  local Hubble flow starting at the Local Group turnaround radius. 
}
\label{fig:radius_velocity_LG}
\end{figure}

\subsection{Systemic Motion and Membership}

The  systemic motion  ($v_{\rm los}$, $\mu_{\alpha\star}$, $\mu_\delta$) of dwarf galaxies and star clusters is key for membership, associations, and to explore the full phase-space.
Figure~\ref{fig:radius_velocity} shows the distance versus velocity for  MW and M31 dwarf galaxies. 
The total velocity can be computed for the majority of the MW dwarf galaxies as systemic proper motions  have been measured with  {\it Gaia} astrometry \citep[e.g.,][]{Simon2018ApJ...863...89S, Fritz2018A&A...619A.103F, McConnachie2020AJ....160..124M,  Battaglia2022A&A...657A..54B, Pace2022ApJ...940..136P}, whereas  systemic proper motion measurements  exist for only a handful of systems from Hubble Space Telescope (HST) astrometry for  M31 dwarf galaxies \citep[e.g.,][]{Brunthaler2007A&A...462..101B, Sohn2020ApJ...901...43S, Warfield2023MNRAS.519.1189W}.
For the MW, I include examples with both the total velocity  and the corrected  Galactocentric radial velocity. 
The corrected velocity  includes a factor of $\sqrt{3}$ to accounts for  the `unknown' tangential velocity and it is included for the MW to show  a pre-{\it Gaia} version. This is a reasonable approximation for the MW and this correction has to be used for almost all other MW-like systems (M31 is the only exception).
For some distant MW systems with large proper motion errors, the tangential velocity may be biased to larger values beyond the MW escape velocity \citep[see discussion in][]{vanderMarel2008ApJ...678..187V, CorreaMagnus2022MNRAS.511.2610C}. 
The last panel of Figure~\ref{fig:radius_velocity} shows  the distance  versus corrected radial velocity relative to M31 for the M31 dwarf galaxy system.  
I include escape velocity curves in Figure~\ref{fig:radius_velocity} to estimate membership in MW or M31 system.
Full orbit modeling is required to confirm membership and  accurate orbital analysis of the MW or M31 needs to  the LMC potential \citep[e.g.,][]{Erkal2020MNRAS.495.2554E, Patel2020ApJ...893..121P, Pace2022ApJ...940..136P, Vasiliev2023Galax..11...59V} or  M33 potential \citep[e.g.,][]{Patel2017MNRAS.464.3825P, Patel2023ApJ...948..104P}, respectively.
The LVDB does not include  orbital modeling results for any system but includes the required input measurements.

Figure~\ref{fig:radius_velocity_LG} shows the distance from the Local Group center versus the corrected velocity relative to the Local Group.
The Local Group coordinates are computed following \citet{Karachentsev1996AJ....111..794K, Karachentsev2002A&A...389..812K} and I assume that the masses of the MW and M31 are equivalent when computing the Local Group coordinates. 
I include escape velocity curves for a point mass of $5\times10^{12} M_{\odot}$ and the local Hubble flow   at the zero-velocity surface located at $R_{LG}\sim1~{\rm Mpc}$ \citep{LyndenBell1981Obs...101..111L, Karachentsev2009MNRAS.393.1265K, Penarrubia2014MNRAS.443.2204P}. 
Galaxies within $R_{LG}$ are bound to the Local Group whereas galaxies beyond $R_{LG}$ follow the local Hubble flow \citep{Karachentsev2009MNRAS.393.1265K}.
A handful of systems have systemic proper motion measurements in the Local Group  \citep[e.g.,][]{Sohn2020ApJ...901...43S, McConnachie2021MNRAS.501.2363M, Battaglia2022A&A...657A..54B, Bennet2024ApJ...971...98B} and additional measurements will be possible in the future with long baselines on current or upcoming space based telescopes \citep[e.g.,][]{vanderMarel2014ASPC..480...43V, Kallivayalil2015arXiv150301785K}.
The velocity relative to the MW, M31, and  Local Group are included in the LVDB catalogs as value-added columns.

\begin{figure*}
\includegraphics[width=\columnwidth]{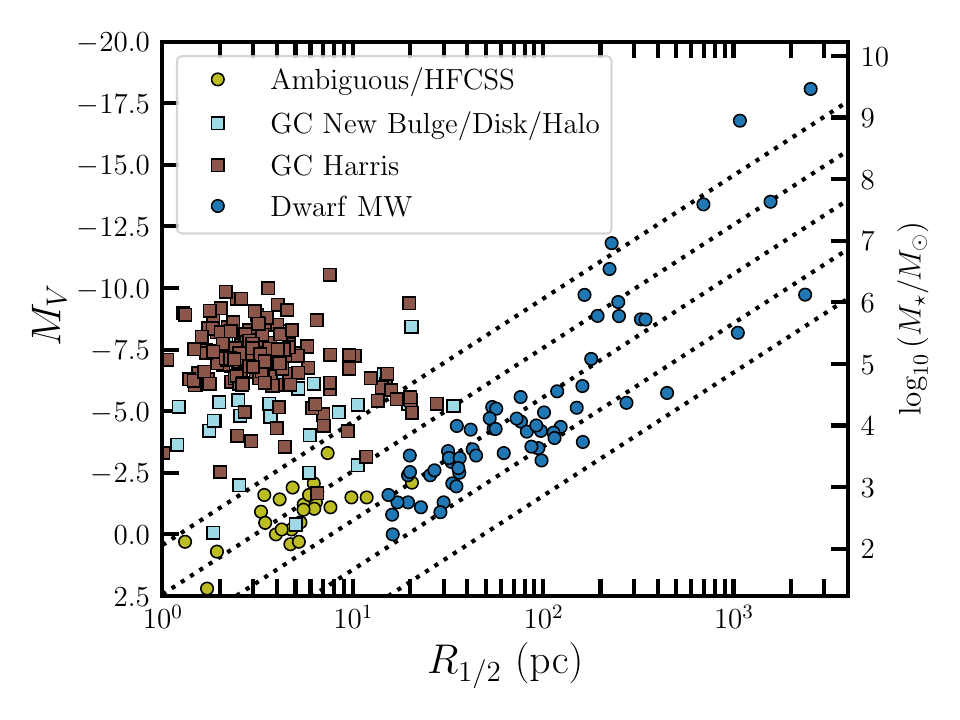}
\includegraphics[width=\columnwidth]{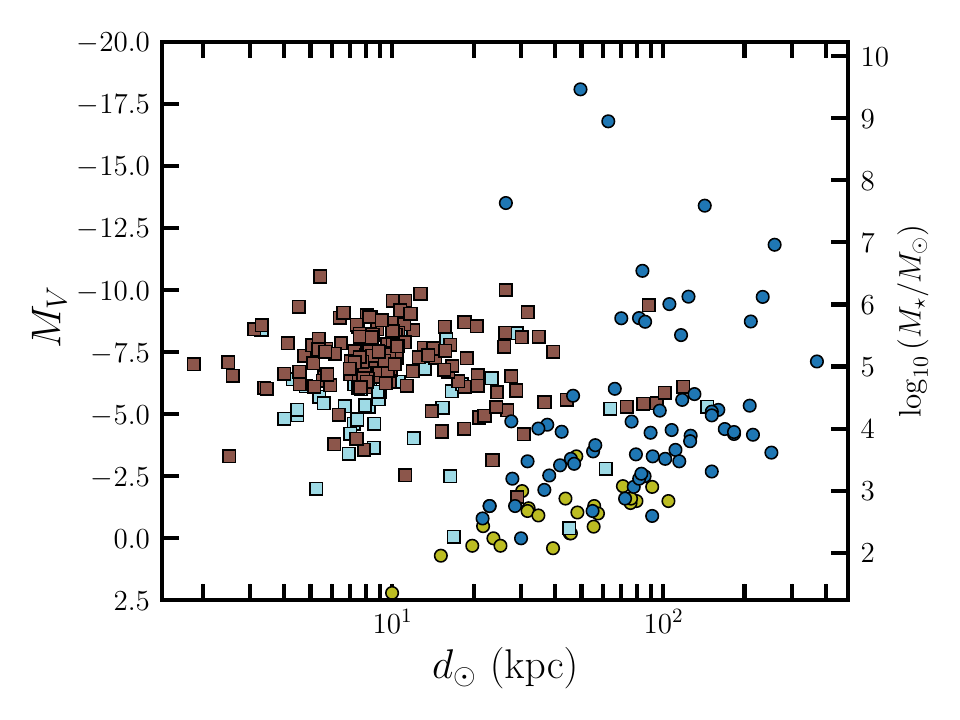}
\caption{{\bf Left}: Azimuthally-averaged  half-light radius ($R_{1/2}$) versus absolute V-band magnitude ($M_V$) and logarithmic stellar mass ($M_\star$) assuming a mass-to-light ratio of 2 for the MW population of dwarf galaxies (blue), classical globular clusters (brown), new or candidate globular clusters  (light blue), and ambiguous or hyper-faint compact stellar systems (olive). {\bf Right}: Heliocentric distance ($d$) versus absolute V-band magnitude  and logarithmic stellar mass  for the same systems.
Contours of  constant surface brightness are indicated with dotted lines at $24,26,28,30,32~{\rm mag~arcsec^{-2}}$.
}
\label{fig:mw_rhalf_distance_mv}
\end{figure*}

\subsection{Structure and Luminosity}

\begin{figure}
\includegraphics[width=\columnwidth]{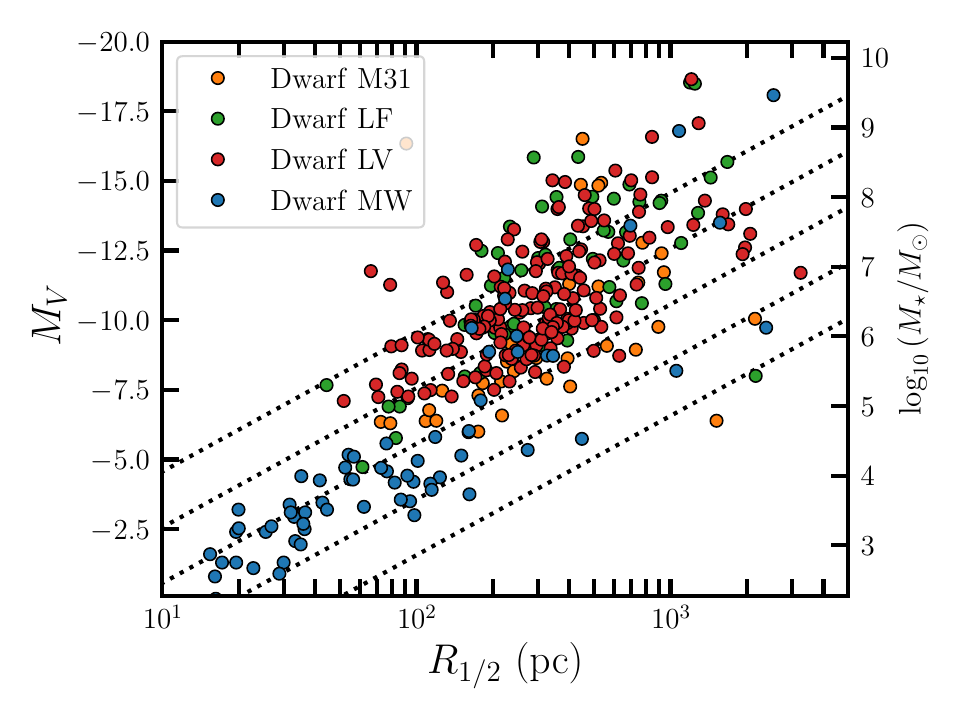}
\caption{Azimuthally-averaged  half-light radius ($R_{1/2}$) versus absolute V-band magnitude ($M_V$) and logarithmic stellar mass ($M_\star$)  for  MW (blue), M31 (orange), Local Field (green), and Local Volume (red) dwarf galaxies.
}
\label{fig:dwarf_rhalf_distance_mv}
\end{figure}

Figure~\ref{fig:mw_rhalf_distance_mv} shows the  size-luminosity plane and distance-luminosity plane of  dwarf galaxies and star clusters  in the MW system. 
For the characteristic size of each system, I utilize the azimuthally-averaged half-light radius (also known as the geometric mean), defined as $R_{1/2} = R_h \sqrt{1-\epsilon}$, where $R_h$ is the half-light radius along the major axis and $\epsilon$ is the ellipticity.  Both latter quantities are compiled as input to the LVDB and $R_{1/2}$ is included as a value-added column.  I use the azimuthally-averaged half-light radius to approximate spherical symmetry as most dwarf galaxies  are non-spherical.
Prior to the discovery of the ultra-faint dwarf galaxies, there was a clear difference in size between globular clusters and dwarf galaxies  with dwarf galaxies above $R_{1/2}\gtrsim 100~{\rm pc}$ and globular clusters below $R_{1/2}\lesssim 20~{\rm pc}$. 
With over 50 new dwarf galaxies,  20  new globular clusters, and  20 HFCSS, there is now overlap in the size-luminosity plane between different types of systems, especially for $R_{1/2} \lesssim 30~{\rm pc}$ and $M_V\gtrsim -5$. 
The region with $M_V > -4$ and $15<R_{1/2}<25~{\rm pc}$ has been referred to as the ``Trough of Uncertainty''  due to the overlap of star cluster and dwarf galaxies and the difficultly of classifying satellites in this region \citep{Conn2018ApJ...852...68C}.

Figure~\ref{fig:dwarf_rhalf_distance_mv} shows the size-luminosity plane for the dwarf galaxy population in the Local Volume.
In general, there is broad agreement in size-luminosity distribution of MW, M31,  Local Field, and Local Volume dwarf galaxies with the caveat that the closer satellite populations  contain fainter dwarf galaxies. 
There are only a few known dwarf galaxies in the ultra-faint region  outside  the MW. This includes $\sim 10$ M31 dwarf galaxies and $\sim5$ galaxies in the Local Field \citep[Leo K, Leo M, Peg W, Tuc B;][]{Sand2022ApJ...935L..17S, McQuinn2023ApJ...944...14M, McQuinn2024ApJ...967..161M}. The ultra-faint dwarf galaxy region is a promising area of discovery space in the upcoming Rubin/LSST, Euclid, and Roman era for more distant MW-like galaxies \citep[e.g.,][]{MutluPakdil2021ApJ...918...88M, Hunt2025A&A...697A...9H}.  

Structural parameter measurements for nearby systems are commonly determined with individual stars \citep[e.g.,][]{Martin2008ApJ...684.1075M, Bechtol2015ApJ...807...50B} whereas the structural parameters of more distant galaxies are commonly determined with integrated light \citep[e.g.,][]{Muller2017A&A...597A...7M}.  Though  space based imaging (e.g., HST) can   resolve  stars  of the distant dwarf galaxies in the Local Volume \citep[e.g.,][]{MutluPakdil2024ApJ...966..188M} and analyze the same type of data to determine structural parameters, it has not been done for many systems. 
Given the different methods, there may be some  systematics in the  structural parameters between  near (resolved stars) and far (integrated light) dwarf galaxies in the Local Volume.
Many studies (especially discovery analyses) utilize single parameter density profiles (e.g., Plummer and Exponential profiles) and more luminous systems have been found to be better fit with two-parameter profiles  \citep[e.g.,][]{Munoz2018ApJ...860...66M}. 
While multiple structural profile parameterizations  can be included in the LVDB, most systems only have a single parameterization included.

\begin{figure*}
\begin{center}
\includegraphics[width=0.9\textwidth]{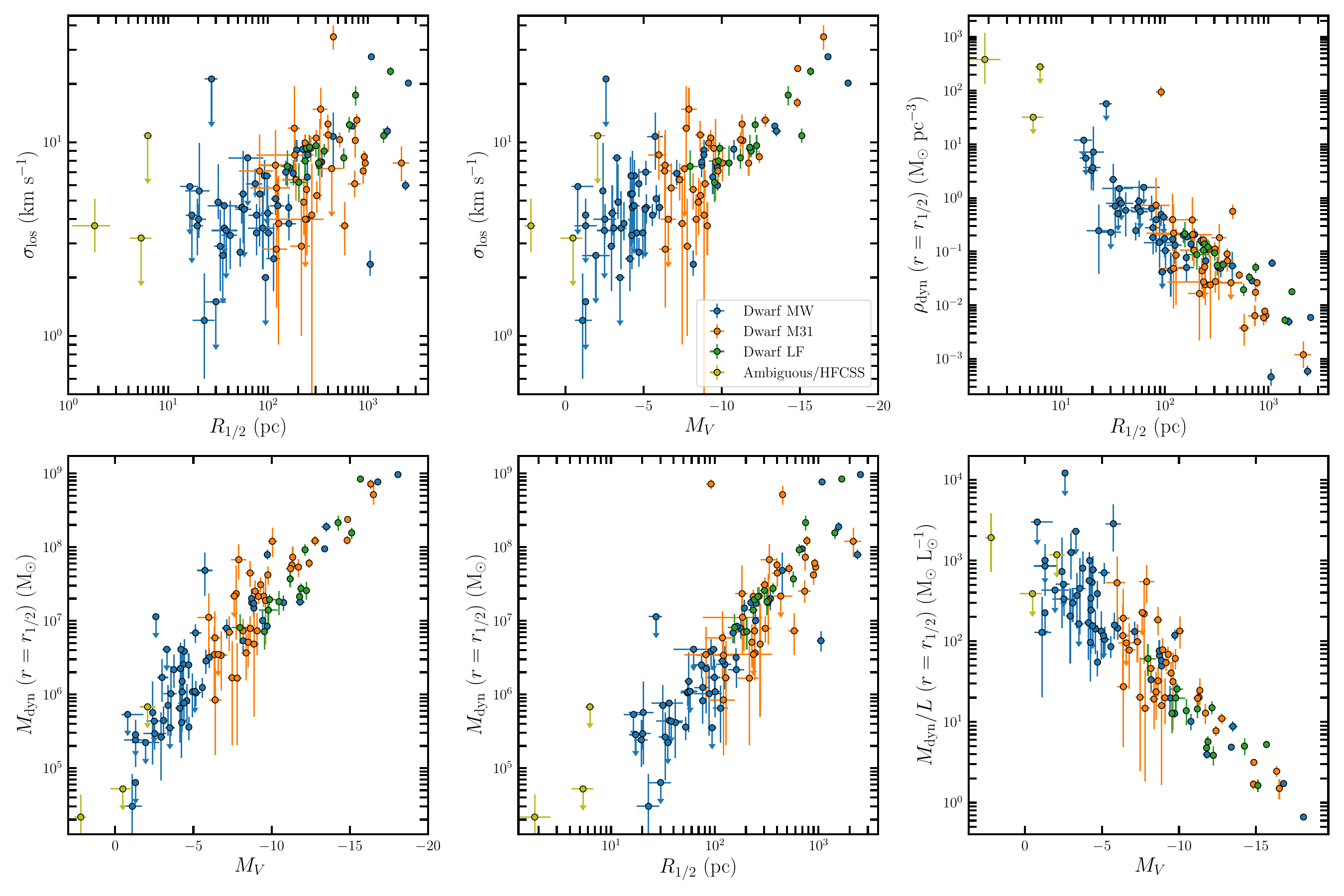}
\caption{{The stellar dynamics  of nearby dwarf galaxies. \bf Top-left}: stellar velocity dispersion ($\sigma_{\rm los}$) versus  projected azimuthally-averaged  half-light radius ($R_{1/2}$) for MW dwarf galaxies (blue), M31 dwarf galaxies (orange),  Local Field dwarf galaxies (green), and ambiguous or HFCSS  (olive).  All panels show the same systems. 
{\bf Top-middle}: stellar velocity dispersion versus absolute V-band magnitude ($M_V$). {\bf Top-right}: Average density within the 3D half-light radius ($\rho_{\rm dym}(r=r_{1/2})$) versus projected azimuthally-averaged half-light radius ($R_{1/2}$).
{\bf Bottom-left}: Dynamical mass within the 3D half-light radius ($M_{\rm dym}(r=r_{1/2})$) versus absolute V-band magnitude ($M_V$).
{\bf Bottom-middle}: Dynamical mass within the 3D half-light radius ($M_{\rm dym}(r=r_{1/2})$) versus azimuthally-averaged projected half-light radius ($R_{1/2}$).
{\bf Bottom-right}: Dynamical mass-to-light ratio within the 3D half-light radius  versus absolute V-band magnitude ($M_V$).
Note that dynamical mass measurements are made within the 3D half-light radius ($r_{1/2}$) and are compared to the 2D projected half-light radius ($R_{1/2}$). For commonly used  stellar density profiles, $r_{1/2}\sim 4/3~R_{1/2}$ \citep{Wolf2010MNRAS.406.1220W}.
}
\label{fig:stellar_kinematics}
\end{center}
\end{figure*}

\subsection{Stellar  Dynamics}

The stellar  dynamical properties of dwarf galaxies in the Local Field are summarized in Figure~\ref{fig:stellar_kinematics}. 
The top-left and top-middle panels of Figure~\ref{fig:stellar_kinematics} compare the velocity dispersion to the stellar size  and luminosity. 
For a pressure-supported system (i.e., a dwarf spheroidal galaxy), the dynamical mass can be computed from the stellar velocity dispersion  \citep[e.g.,][]{Battaglia2022NatAs...6..659B}. While no results from detailed dynamical models are included in the LVDB, the estimated dynamical mass ($M_{1/2}=M(r=r_{1/2})$) within the 3D deprojected half-light radius\footnote{For many commonly used stellar profiles, the 3D deprojected half-light radius  is roughly proportional to the 2D projected half-light radius, $r_{1/2}\sim 4/3 R_{1/2}$ \citep{Wolf2010MNRAS.406.1220W}.} ($r_{1/2}$) is computed from  the velocity dispersion using the \citet{Wolf2010MNRAS.406.1220W} mass estimator\footnote{See  \citet{Walker2009ApJ...704.1274W, Errani2018MNRAS.481.5073E} for other mass estimators.}.
The dynamical mass  is compared to the stellar size and  luminosity in bottom-left and bottom-middle  panels of Figure~\ref{fig:stellar_kinematics}.
The top-right  and  bottom-right panels compare the average density within the 3D half-light radius   and the dynamical mass-to-light ratio to  luminosity. 
The dwarf galaxies of the Local Group are among the most dark matter dominated systems in the local universe and are excellent tracers to probe the nature of dark matter \citep[e.g.,][]{Gilmore2007ApJ...663..948G, Chen2017MNRAS.468.1338C, Valli2018NatAs...2..907V, Kim2021arXiv210609050K, Hayashi2021PhRvD.103b3017H, Dalal2022PhRvD.106f3517D}.

The internal dynamics of dwarf galaxies and star clusters of nearby systems ($d\lesssim 1 ~{\rm Mpc}$) are  generally measured with radial velocities of individual stars. Due to the faintness of individual stars, this requires multi-object spectroscopy with 8-m class telescopes. Sample sizes range from a several hundreds to a couple thousand in classical satellites \citep[e.g.,][]{Walker2009AJ....137.3100W, Fabrizio2016ApJ...830..126F, Pace2020MNRAS.495.3022P, Walker2023ApJS..268...19W} and from  tens to almost a hundred stars in ultra-faint dwarf galaxies \citep[e.g.,][]{Simon2007ApJ...670..313S, Martin2007MNRAS.380..281M, Koposov2015ApJ...811...62K, Jenkins2021ApJ...920...92J, Walker2023ApJS..268...19W, Heiger2024ApJ...961..234H}.
These studies commonly target either the Calcium triplet \citep[e.g.,][]{Tolstoy2023A&A...675A..49T} or Magnesium triplet \citep[e.g.,][]{Walker2023ApJS..268...19W} absorption features for precise and accurate radial velocities.
While  radial velocities of individual star can be  obtained for some nearby Local Field galaxies \citep[e.g.,][]{Kirby2014MNRAS.439.1015K}, going to larger distances requires utilizing Integral Field Unit spectroscopy \citep[e.g., VLT/MUSE, KECK/KCWI;][]{Danieli2019ApJ...874L..12D, Fahrion2020A&A...634A..53F, Muller2021A&A...645A..92M}, the James Webb Space Telescope  \citep[JWST; e.g.,][]{Nidever2024IAUS..377..115N}, or future 30m class telescopes.

For ultra-faint dwarf galaxies, one difficult systematic is unresolved  binary stars. Systems with the smallest intrinsic velocity dispersion ($\sigma_{\rm los}\lesssim 5~{\rm km~s^{-1}}$), with small spectroscopic sample sizes ($N\lesssim 10-20$), and/or without multi-epoch spectroscopic data can have incorrect inferred velocity dispersions if there are unidentified binary stars included.
Multiple ultra-faint dwarf galaxies have had their inferred velocity dispersion and dynamical mass reduced after multi-epoch data  identified binary stars \citep[e.g., Tri~II, Gru~I, Her, Tuc~II, Boo~II;][]{Kirby2017ApJ...838...83K, Chiti2022ApJ...939...41C, Ou2024ApJ...966...33O, Chiti2023AJ....165...55C, Bruce2023ApJ...950..167B}. I suspect that some ultra-faint dwarf galaxies with the highest dynamical mass-to-light ratios may contain unidentified  binary stars in their current spectroscopic samples and may have inflated dynamical masses. 

The stellar kinematics in the LVDB are primarily characterized by the velocity dispersion describing pressure-supported systems.
There are some nearby dwarf spheroidal galaxies that are clearly rotating  \citep[e.g., And~II, Phoenix;][]{Ho2012ApJ...758..124H, Kacharov2017MNRAS.466.2006K} and stellar rotation is not included in the dynamical mass measurements\footnote{One suggested correction is to replace the  velocity dispersion in the mass estimator with a combination of the velocity dispersion and rotation velocity $\sigma_{\rm los}^2 \to \sigma_{\rm los}^2 + (v_{\rm rot}\sin{i})^2$ \citep{Weiner2006ApJ...653.1027W, Kirby2014MNRAS.439.1015K}} in Figure~\ref{fig:stellar_kinematics}.

\subsection{Stellar Metallicity}
\label{section:metallicity}

\begin{figure*}
\begin{center}
\includegraphics[width=0.9\textwidth]{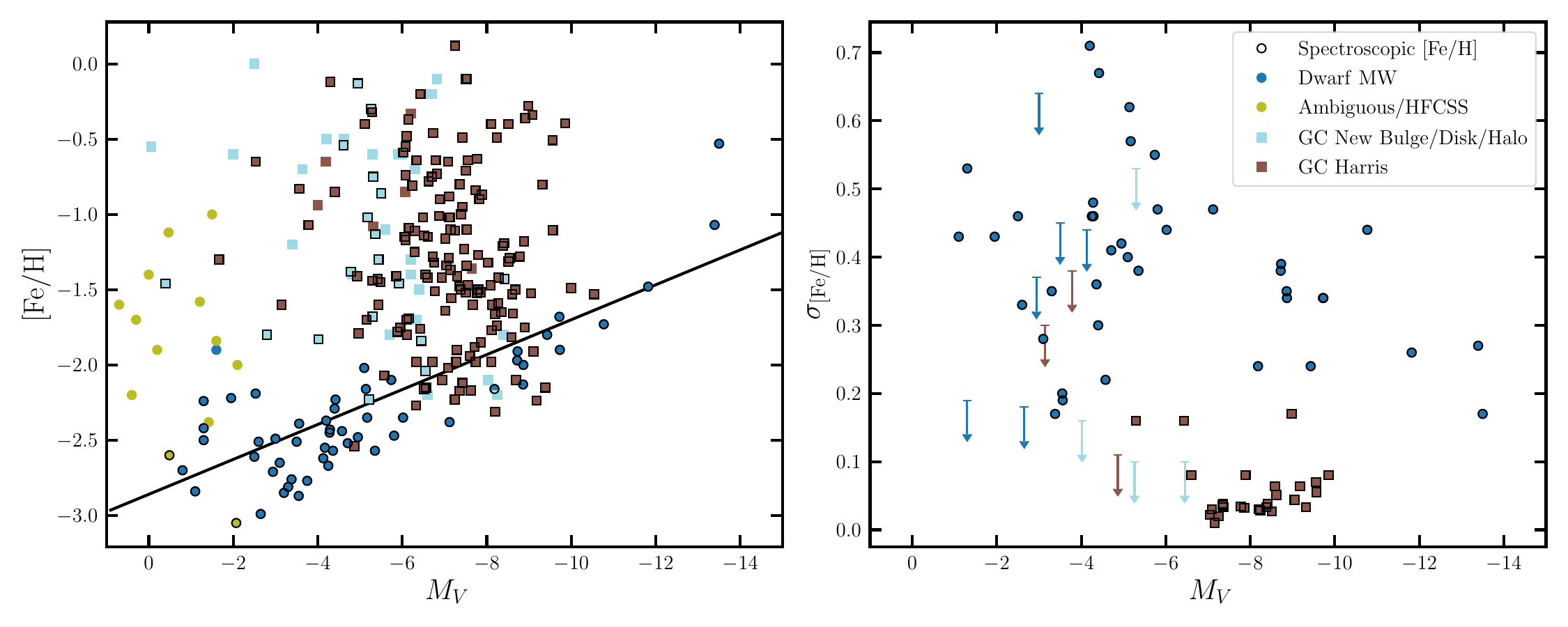}
\caption{{\bf Left}: Absolute V-band magnitude ($M_V$) versus mean stellar metallicity  ([Fe/H]) for satellites   in the MW system including:  MW dwarf galaxies (blue), ambiguous or hyper-faint compact stellar systems (olive), new  globular clusters in the bulge, disk, and halo (light blue), and globular clusters in the Harris catalog (brown). Systems with spectroscopic metallicities are outlined in black circles and the stellar mass-stellar metallicity relation from \citet{Simon2019ARA&A..57..375S} is included.   {\bf Right}: Absolute V-band magnitude versus metallicity dispersion ($\sigma_{\rm [Fe/H]}$) for the same systems. 
}
\label{fig:metallicity_mw}
\end{center}
\end{figure*}

The mean stellar metallicity ([Fe/H]) and metallicity dispersion of dwarf galaxies  and globular clusters  in the MW system is shown in Figure~\ref{fig:metallicity_mw}. The LVDB contains metallicity measurements from  photometry (e.g.,  metallicity sensitive   photometry), isochrone/color-magnitude diagram fits, and spectroscopic metallicity (either from high or medium resolution spectroscopy). 
Spectroscopic metallicities are favored over isochrone and photometric metallicity\footnote{I note that in some cases, much larger samples of metallicity measurements for individual stars    can be obtained with metallicity sensitive narrowband photometry (e.g., Ca H\&K)  compared to spectroscopic samples \citep[e.g.,][]{Fu2023ApJ...958..167F, Fu2024ApJ...965...36F} and these samples would better describe a system's metallicity distribution than  a small spectroscopic sample. } for summary measurements in the LVDB and I outline   systems with spectroscopic metallicity in Figure~\ref{fig:metallicity_mw}. 
While the MW dwarf galaxies display a clear trend with stellar mass and metallicity \citep[i.e., they follow the stellar mass-metallicity relation; ][]{Kirby2013ApJ...779..102K}, there is no obvious trend for the MW globular cluster system.
The metallicity dispersion is a useful measurement to distinguish dwarf galaxies from star clusters as dwarf galaxies will have a metallicity  spread while star clusters\footnote{I note that the LVDB is generally incomplete for existing   metallicity dispersion measurements for globular clusters.} will generally   not.
Along these lines, there is an observed   metallicity floor at ${\rm [Fe/H]}\sim -2.5$ in the MW globular clusters system \footnote{Although there are two MW globular cluster stellar streams and an M31 globular cluster below the MW metallicity floor \citep{Wan2020Natur.583..768W, Larsen2020Sci...370..970L, Martin2022Natur.601...45M}.} 
and the  mean stellar metallicity can be used with the the stellar mass-metallicity relation   to assist with classification \citep[e.g.,][]{Simon2017ApJ...838...11S}. 

\begin{figure*}
\begin{center}
\includegraphics[width=0.9\textwidth]{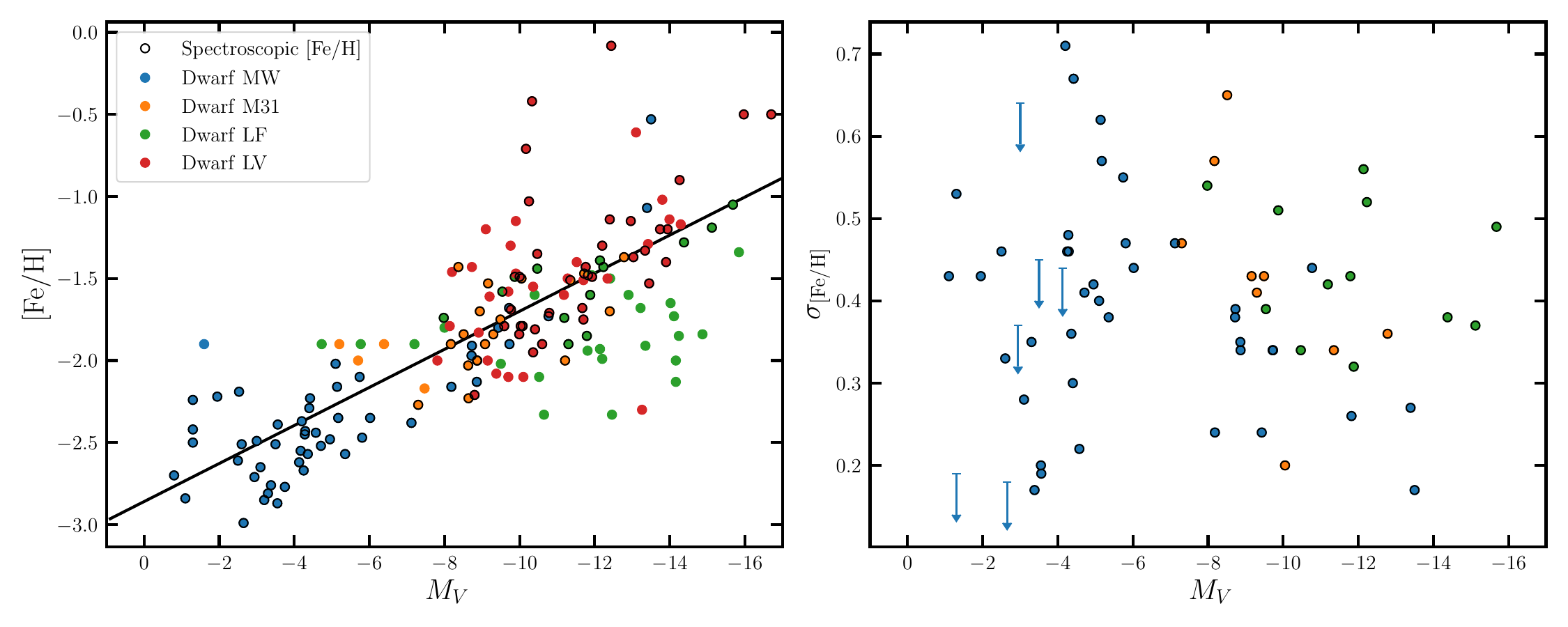}
\caption{{\bf Left}: Absolute V-band magnitude ($M_V$) versus mean stellar metallicity  ([Fe/H])  for MW  (blue), M31 (orange), Local Field  (green), and Local Volume  (red) dwarf galaxies.  Systems with spectroscopic metallicities are outlined in black circles and the stellar mass-stellar metallicity relation from \citet{Simon2019ARA&A..57..375S} is included.  {\bf Right}: Absolute V-band magnitude  versus metallicity dispersion ($\sigma_{\rm [Fe/H]}$) for the same systems. 
}
\label{fig:metallicity_dwarf}
\end{center}
\end{figure*}

Figure~\ref{fig:metallicity_dwarf} shows the summary of the mean metallicity and metallicity dispersion for the dwarf galaxies in the Local Volume. 
For dwarf galaxies in the Local Volume, there is a clear trend between stellar mass and stellar metallicity \citep{Kirby2013ApJ...779..102K}.
The stellar mass-stellar metallicity relation and correlation is used to test chemical evolution models, inflows and outflows, and stellar feedback models \citep[e.g.,][]{Garnett2002ApJ...581.1019G, Finlator2008MNRAS.385.2181F, Kravtsov2022MNRAS.514.2667K}.
There is an observed metallicity plateau at ${\rm [F/H]}\sim-2.6$ for the faintest MW dwarf galaxies \citep{Fu2023ApJ...958..167F}.
The inclusion of the intergalactic medium metallicity in semi-analytic models can explain the observed metallicity floor \citep{Kravtsov2022MNRAS.514.2667K, Ahvazi2024MNRAS.529.3387A}.
The metallicity dispersion for dwarf galaxies ranges from 0.2-0.7 dex and there is no clear trend with luminosity. 
Less work has been done to explain the large range in metallicity dispersion and physical processes include the stochasticity of chemical enrichment processes and (inhomogeneous) metal mixing \citep[e.g.,][]{Greif2010ApJ...716..510G, Frebel2014ApJ...786...74F, Emerick2020ApJ...890..155E}.
Understanding the stellar metallicity distribution of dwarf galaxies in the Local Group is key for testing stellar feedback and chemical evolution models.

\begin{figure*}
\begin{center}
\includegraphics[width=0.9\textwidth]{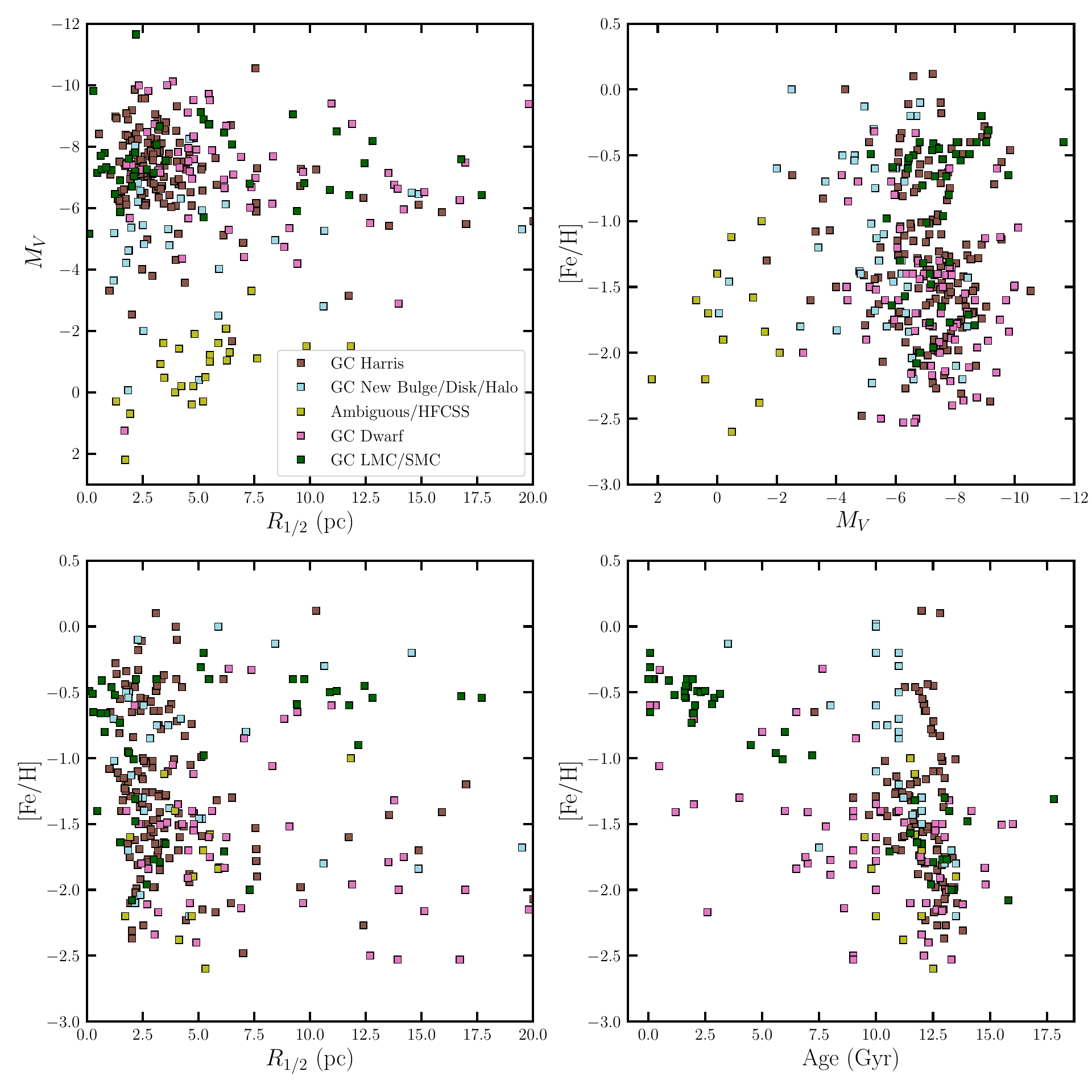}
\caption{Properties of globular clusters and star clusters in the MW system. The systems are the same for all panels and include:   classical/\citet{Harris1996AJ....112.1487H} catalog globular clusters (brown), new globular clusters and candidates in Galactic disk, bulge, and halo (light blue), ambiguous or hyper-faint compact stellar systems (olive), star clusters in dwarf galaxies (pink), and LMC/SMC star clusters (dark green).  {\bf Top-left}: azimuthally-averaged  half-light radii  ($R_{1/2}$) versus  absolute V-band magnitude ($M_V$).
{\bf Top-right}:  absolute V-band magnitude ($M_V$) versus mean metallicity  ([Fe/H]).
{\bf Bottom-left}: azimuthally-averaged  half-light radii  ($R_{1/2}$) versus mean metallicity ([Fe/H]).
{\bf Bottom-right}: age versus mean metallicity ([Fe/H]).
}
\label{fig:gc_summary}
\end{center}
\end{figure*}

\subsection{Globular Clusters}

The LVDB  has several  globular clusters and star clusters catalogs including:  Harris/classical, new bulge/disk/halo, dwarf galaxy hosted, and the ambiguous hyper-faint compact  stellar systems.
The majority of the globular clusters in the LVDB catalogs are located in MW or its satellites. 
The only globular clusters beyond the MW are hosted in nearby Local Field or Local Volume dwarf galaxies. 
The properties of the  globular clusters in the LVDB (size, luminosity, metallicity, and age)  are summarized in Figure~\ref{fig:gc_summary}. 
The majority of the globular clusters are bright ($M_V<-5$) but there is a tail to fainter systems. There is a similar tail to larger sizes ($R_{1/2}>10~{\rm pc}$). 
Whether these systems lost their stellar mass due to internal dynamical effects or external tidal forces is an open question and likely depends on properties specific to each individual cluster.

While one motivation for the globular cluster compilations in the LVDB is for  comparisons to  newly discovered stellar systems  and assist in classification, there are  many new  systems that have not been compiled in the literature.
This includes the ambiguous systems or hyper-faint compact stellar systems, globular clusters hosted by dwarf galaxies, and the more recently discovered globular clusters and candidates\footnote{Although  two recent studies have compiled newly discovered globular clusters and candidates in the Galactic bulge \citep{Bica2024A&A...687A.201B, Garro2024A&A...687A.214G}.}. 
More recently discovered globular clusters are generally fainter ($M_V>-5$) and/or located in regions with higher stellar densities (i.e., Galactic bulge and disk). 
While the number of bright globular clusters in the MW halo and at higher Galactic latitude is thought to be roughly complete \citep[e.g.,][]{Webb2021MNRAS.502.4547W} there are many new candidates at low Galactic latitudes  \citep[e.g.,][]{Minniti2011A&A...527A..81M, Gran2022MNRAS.509.4962G}.

At a population level, the kinematics of globular clusters along with their  age and metallicity have been used to understand their origin and   connection to accretion events \citep{Massari2019A&A...630L...4M, Kruijssen2019MNRAS.486.3180K}.
Recent analysis has studied the census of globular clusters and candidates in Galactic bulge  and compared the globular cluster  metallicity distribution to the in-situ MW stellar populations and ex-situ accretion events \citep{Bica2024A&A...687A.201B, Garro2024A&A...687A.214G}.
Properties such as the number and mass in globular cluster systems correlates with the stellar and halo mass of the host galaxy \citep[e.g.,][]{Spitler2009MNRAS.392L...1S, Forbes2018MNRAS.481.5592F, Eadie2022ApJ...926..162E} and the many dwarf galaxies without globular clusters can still be used to determine  these relations \citep[e.g.,][]{Berek2023MNRAS.525.1902B}.
The LVDB compilation of globular  clusters hosted by dwarf galaxies will be useful for studies on understanding the connection between  globular cluster systems and host galaxy properties. 
In particular, why do some low mass dwarf galaxies host globular clusters while other dwarf galaxies at the same stellar mass do not?

\begin{figure}
\includegraphics[width=\columnwidth]{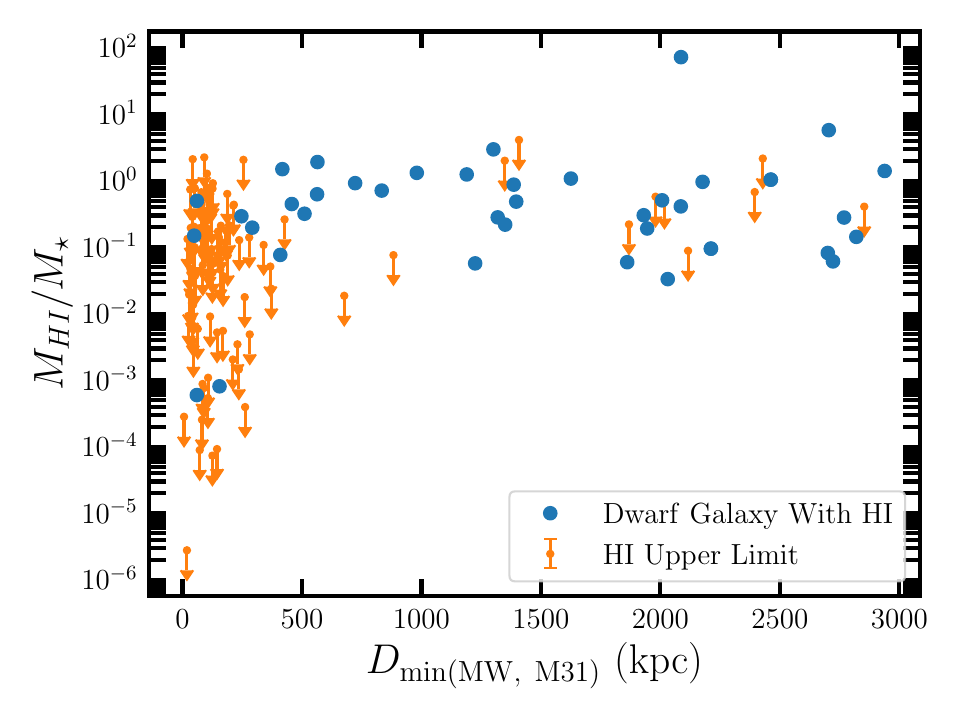}
\caption{Distance to the closest massive galaxy (MW or M31) versus the ratio of HI gas mass to stellar mass versus  for dwarf galaxies in the Local Field. The majority of satellites are quenched with the exception of some   massive satellites    whereas many  isolated dwarf galaxies  retain their gas reservoir.  
}
\label{fig:ratio_gas_stellar}
\end{figure}

\subsection{Neutral Hydrogen Gas}

The LVDB includes measurements of the neutral hydrogen gas flux.
Figure~\ref{fig:ratio_gas_stellar} shows the ratio of HI  mass to stellar mass versus the minimum distance to the MW or M31 for Local Field dwarf galaxies. 
The majority of satellites are quenched except for some   massive satellites     whereas many  isolated dwarf galaxies  retain their gas reservoir \citep[e.g.,][]{Einasto1974Natur.252..111E}.
There are several environmental processes that can strip the HI gas and quench infalling low mass dwarf galaxies such as ram-pressure stripping and turbulent viscous stripping \citep[e.g.,][]{Fillingham2016MNRAS.463.1916F}.
The primary HI property cataloged in the LVDB  is the HI flux. One key  HI property that is missing in the LVDB are   kinematic measurements and their inclusion would  to increase the number of low mass dwarf galaxies with dynamic mass measurements \citep[e.g.,][]{Oh2015AJ....149..180O}. 
A significant number of  systemic velocity measurements of  distant dwarf galaxies  in the LVDB are from HI results.

\subsection{Ambiguous or Hyper-Faint Compact Stellar Systems}
\label{section:ufcss}

\begin{figure*}
\begin{center}
\includegraphics[width=0.9\textwidth]{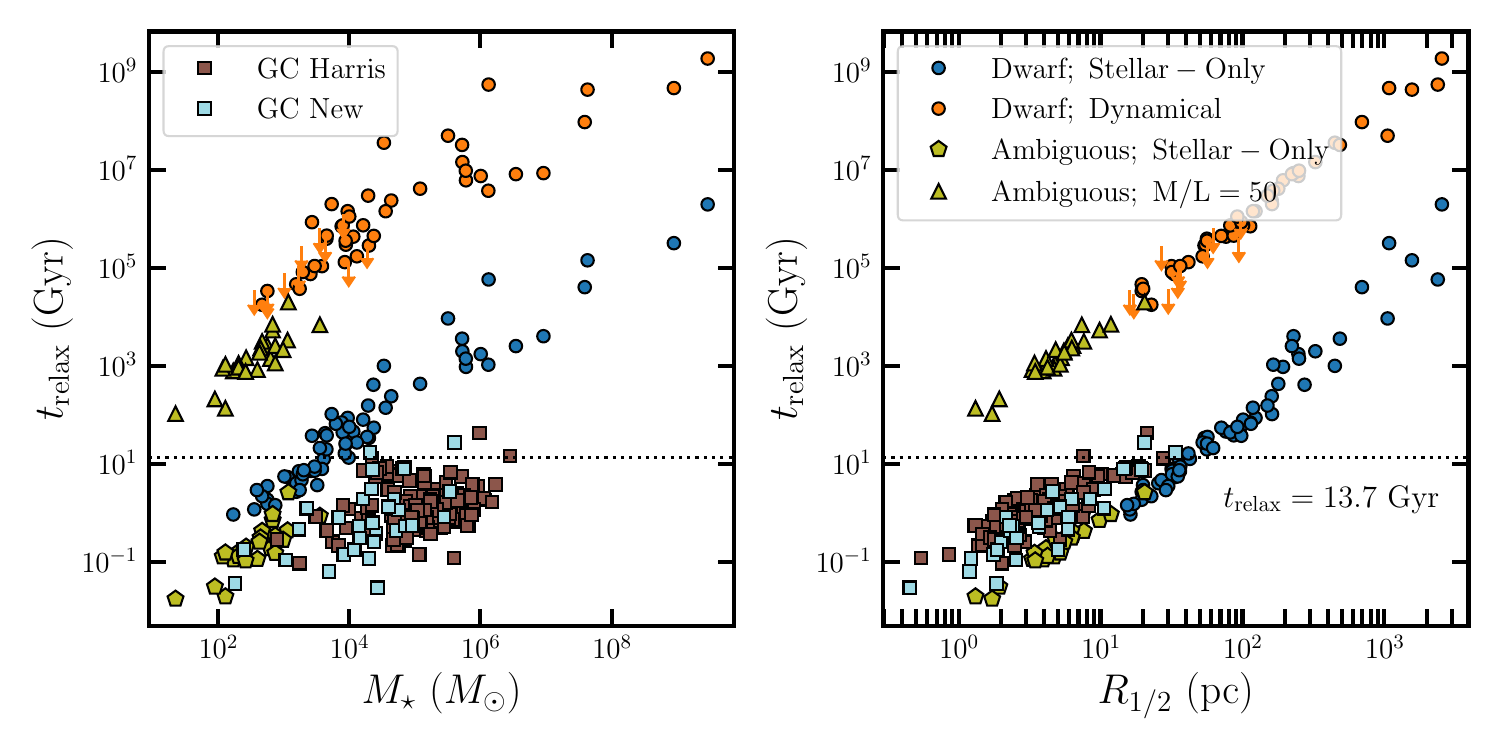}
\caption{ Relaxation time for MW satellites compared to their stellar mass ({\bf left}) and stellar size ({\bf right}). 
For the MW dwarf galaxies and the ambiguous/HFCSS satellites both  stellar-only and dynamical relaxation times are included.  The ambiguous/HFCSS dynamical relaxation times assume $M/L=50$.
A dotted line is included to denote the age of the Universe. Star clusters will have two-body relaxation times lower than the age of the universe whereas dwarf galaxies will have much larger timescales.
}
\label{fig:timescales}
\end{center}
\end{figure*}

The ambiguous or  HFCSS\footnote{I note that these systems have many names in the literature: ultra-faint star clusters, micro galaxies/glofs \citep{Errani2024ApJ...965...20E, Errani2024ApJ...968...89E}, ultra-faint objects \citep{Contenta2017MNRAS.466.1741C}, ultra-faint compact satellites, ultra-faint compact stellar systems \citep{Circiello2025ApJ...978L..43C}, faint ambiguous satellites \citep{Smith2024ApJ...961...92S}, faint halo clusters \citep{Bica2019AJ....157...12B}, hyper-faint compact stellar systems and/or ambiguous systems (this work). I (and some colleagues) have referred to these systems as ultra-faint star clusters for years which incorrectly suggested a classification for these systems.}  are a category of systems compiled in the LVDB for the first time.
One goal of the HFCSS compilation is to encourage the community to study these mostly forgotten systems and assist in classifying them. 
The HFCSS  populate the compact  ($R_{1/2}<10-20~{\rm pc}$) and faint ($M_V> - 3$) region of the size-luminosity plane. 
The HFCSS are located in the region of the size-luminosity plane when both the star cluster or dwarf galaxy populations would be located if  extrapolated   to fainter magnitudes.
This small and faint region has been referred to as the ``valley of ambiguity''  or the ``trough of uncertainty.''
Of note, all known systems are located in the Galactic halo and  could not be found at low Galactic latitudes due to higher stellar backgrounds. 
It is unclear whether these systems have undergone extreme  tidal disruption \citep[e.g.,][]{Errani2024ApJ...968...89E}. 
Many of these systems were initially considered star clusters\footnote{Almost of these systems follow  star cluster naming conventions.} based on photometric analysis but the majority lack any spectroscopic follow-up. 
Compared to when the first HFCSS were discovered, there are  now theoretical predictions that  hyper-faint dwarf galaxies should exist in the MW to $M_V\sim0$ \citep[e.g.,][]{Manwadkar2022MNRAS.516.3944M}.

There are already systems in the dwarf galaxy  or  globular cluster catalogs that would be considered  HFCSS if they lacked spectroscopic follow-up. 
There are four globular clusters in the Harris catalog, AM 4, Palomar 1, Palomar 13, and 2MASS GC-02, that have sizes and luminosity that are in (or just outside) the HFCSS region. 
All have spectroscopic follow-up that suggests they are faint globular clusters 
\citep[e.g.,][]{Bradford2011ApJ...743..167B, Kunder2021AJ....162...86K}.
Similarly, there are several  recently discovered star clusters in the HFCSS region, Laevens~3, Mu{\~n}oz~1, and Segue~3,  with    spectroscopic follow-up that classifies them as star clusters \citep{Fadely2011AJ....142...88F, Munoz2012ApJ...753L..15M, Longeard2019MNRAS.490.1498L}.
There are several dwarf galaxies in the HFCSS region, Draco~II, Triangulum~II, Segue 1, Segue~2, Tucana~V, and Willman~1, with  a metallicity spread and/or a resolved velocity dispersion from spectroscopic follow-up \citep{Kirby2013ApJ...770...16K, Kirby2017ApJ...838...83K, Fu2023ApJ...958..167F, Simon2011ApJ...733...46S, Hansen2024ApJ...968...21H, Fu2023ApJ...958..167F, Willman2011AJ....142..128W}.
Some of these systems may be tidally influenced; for example, Segue~1, Triangulum~II, and Willman~1 have orbital pericenters of 20, 12, and 16~kpc, respectively \citep{Pace2022ApJ...940..136P}.
At least one system, Tucana~III, has no resolved velocity or metallicity dispersion but has been considered a dwarf galaxy due to its size, mean metallicity, orbit, and tidal disruption \citep{Simon2017ApJ...838...11S, Li2018ApJ...866...22L}.
The  Ursa Major III/UNIONS~1  system has a resolved velocity dispersion but the removal of 1-2 potential outlier stars from the measurement leads to an unresolved velocity dispersion and without a metallicity measurement the classification remains unclear \citep{Smith2024ApJ...961...92S}  and I consider it a HFCSS here.
The systems, Draco~II and Eridanus~III, have conflicting classification evidence as there is  observed mass segregation in both systems    that suggests  a star cluster origin \citep{Baumgardt2022MNRAS.510.3531B} while both systems also have a resolved metallicity dispersion from metallicity sensitive narrowband photometry that suggests a dwarf galaxy origin \citep{Fu2023ApJ...958..167F}. The main difference between the two systems is their size and Draco~II is currently classified as a galaxy while Eridanus~III is considered a HFCSS. 
I note that several  HFCSS have evidence for mass segregation: Kim~2, Kim~3 \citep{Kim2015ApJ...803...63K, Kim2016ApJ...820..119K},  Balbinot 1, DES 1,  Koposov 1,  Koposov 2,  and Segue 3 \citep{Baumgardt2022MNRAS.510.3531B}.
Even with spectroscopy, the faintest HFCSS may be difficult to classify. Spectroscopy of DELVE~1 and Eridanus~III has found the brightest star of both systems to be very metal-poor ([Fe/H]$\sim-3.1, -2.8$) and Carbon-enhanced \citep{Simon2024ApJ...976..256S} but lack spectroscopic metallicity spread constraints due to the lack of stars bright enough of  measurements.  Carbon-enhanced stars are extremely rare in star clusters. 

One proposed definition to distinguish star clusters and  galaxies is that  the stars are collisionless in galaxies and have a  two-body relaxation time longer than a Hubble time \citep{Forbes2011PASA...28...77F}. To explore this with the LVDB, in Figure~\ref{fig:timescales} I compare the two-body relaxation time \citep[following,][]{Binney2008gady.book.....B, Errani2024ApJ...965...20E} to the stellar mass and stellar size of MW satellites.  
I compute the relaxation time for the globular clusters based purely  on their stellar properties where as for the dwarf galaxies I compute both a stellar  and  dynamical relaxation time.
For pure stellar systems such as globular clusters, the two-body relaxation time is shorter than a Hubble time. Most dwarf galaxies, even if they were pure stellar systems, would have relaxation times longer than a Hubble time due to their large sizes. When the dynamical mass is considered for the relaxation time, all dwarf galaxies have a relaxation time larger than the age of the universe.  Due to their small sizes and stellar masses, all the HFCSS have small relaxation times from their stellar components.  If I instead assume they had larger mass-to-light ratios of 50  their dynamical relaxation times are significantly larger \citep[see ][ for a more in depth discussion]{Errani2025arXiv250522717E}.  The small stellar-only relaxation times show why these systems were almost all initially considered stars clusters. 
As  the dynamical relaxation time and dynamical mass-to-light ratio both require stellar kinematic measurements, both tests will have  similar classification results. 
Searching for mass segregation is an alternative method to infer short relaxation times and assist with classification of the HFCSS.

In summary, the MW  contains a number of compact ($R_{1/2}<20~{\rm pc}$) and faint ($M_V>-3$) satellites. While some systems with spectroscopic follow-up in this size and luminosity range have been confidently classified as star clusters or dwarf galaxies, the majority lack any spectroscopic follow-up and cannot be classified as dwarf galaxies or star clusters.  These systems represent either the extreme faint end of the star cluster population or the extreme faint end of the dwarf galaxy population. Regardless of their classification, the systems may have undergone extreme tidal stripping and they may have tidal tails below current surface brightness limits. 
The existence of the HFCSS can help identify the minimum halo mass for galaxy formation and/or help understand how some systems survive in tidal fields.

\section{Issues, Limitations, Expansion, and Future Development}
\label{section:issues}

There are some current issues and limitations with the LVDB along with many ideas or requests for expanding the LVDB.
The current format of the LVDB does not limit the resolution of these issues and enables the expansion of the LVDB for  additional properties and measurements. Community feedback, assistance, and contribution will drive the resolutions of these issues  and is welcome.

\begin{itemize}
\item The current structure of the database was created and optimized for studies of MW dwarf spheroidal galaxies. These systems are all resolved into individual stars and excluding the LMC and SMC,  lack HI gas. 
\item The database includes a variety of techniques and methods for both  measurements and analyses. Due to this inhomogeneity, there will be systematics in the LVDB and some studies may opt for homogenized measurements.
\item As previously mentioned, many catalogs are currently incomplete for \textit{known} systems in the Local Volume. This includes  dwarf galaxies beyond $3~{\rm Mpc}$, LMC/SMC star clusters, and  star clusters beyond the MW including dwarf galaxy hosted star clusters. 
In addition, there are several obvious types of systems such as the  M31 globular clusters\footnote{There are two existing catalogs that summarize the properties of the M31 globular clusters: ``The Revised Bologna Catalogue''  \citep[\url{http://www.bo.astro.it/M31/}][]{Galleti2004A&A...416..917G}  and ``Studies of Resolved Objects in M31'' \url{https://lweb.cfa.harvard.edu/oir/eg/m31clusters/M31_Hectospec.html}. } and the MW open clusters\footnote{There are several existing  catalogs that summarize the census and properties of MW open clusters  \citep[e.g.,][]{Bica2019AJ....157...12B, CantatGaudin2020A&A...633A..99C, Hunt2023A&A...673A.114H}.} that are entirety absent from the LVDB. 
\item When new \texttt{YAML} collections and keys have been added, there has not been a significant effort to populate measurements for existing systems  (e.g., type, E(B-V), star formation history/quenching times). 
\item Many systems are missing discovery references. This is especially true for systems discovered before the start of my academic career.
\item Some galaxies  show changes in the position angle and/or ellipticity with radius and the current LVDB format does not account for this.  
\item The current stellar kinematic setup assumes a system has a constant velocity dispersion and no rotation. While this has been traditionally  true for MW dwarf spheroidal galaxies \citep[e.g.,][]{Walker2007ApJ...667L..53W}, some  dwarf spheroidal galaxies are  clearly rotating \citep[e.g., And~II;][]{Ho2012ApJ...758..124H} and others  may contain velocity gradients \citep[e.g., Leo~V, Ant~II;][]{Collins2017MNRAS.467..573C, Ji2021ApJ...921...32J}.
\item The stellar metallicity distribution is based on the mean (or median) and dispersion. Most observed metallicity distributions are generally non-Gaussian and skewed \citep[e.g.,][]{Kirby2011ApJ...727...78K} and most dwarf galaxies have radial metallicity gradients \citep[e.g.,][]{Taibi2022A&A...665A..92T}. 
\item There is existing HI kinematic literature that is not included for many galaxies and there are other  HI summary properties that could be included (line width profile measurements or the flat velocity of the rotation curve, $v_{\rm flat}$).
\item There are multiple forms of the Sersic profile in the literature  with some studies parameterizing the profile in terms of a generic scale radius whereas others use the  effective radius.  While an effort has been made to homogenize these measurements, there may be some inconsistencies.
\item There are a number of systems without an independent distance measurement and they inherit the  distance and errors of their host galaxy.  This is generally globular clusters hosted by dwarf galaxies and distant dwarf galaxy candidates.  This is not denoted in the combined catalogs. 
\item The LVDB currently only includes $V$-band magnitude measurements (both apparent and absolute) and  it would be beneficial to include  magnitudes in both other bands and other photometric systems (UBVI, SDSS, etc). 
\item The stellar kinematics are based solely of the line-of-sight kinematics. The  kinematics from internal tangential motions (i.e., proper motions) are available for many of the closer  MW globular clusters  from {\it Gaia} or HST astrometry \citep[e.g.,][]{Vasiliev2021MNRAS.505.5978V, Libralato2022ApJ...934..150L} and several dwarf galaxies  \citep{Massari2018NatAs...2..156M, Vitral2024ApJ...970....1V}. Internal tangential kinematic measurements will be enabled with future long-term astrometric measurements.
\item The central surface brightness is not included. 
\item I have chosen not to include orbital information (pericenter, apocenter, etc). These quantities are  dependent on a number of model choices such as the MW potential,  LMC inclusion, and/or solar motion. 
\item There are a number of other observational properties that could be included: RR Lyrae or other variable star summary statistics,  gas phase metallicity (O/H), and star formation rates tracers such as the flux in $H_{\alpha}$, near- and far-UV.
\end{itemize}

\section{Summary}
\label{sec:conclusion}

The Local Volume database (LVDB) is a compilation of observed properties of dwarf galaxies and star clusters in the Local Volume. 
I have presented an overview of the structure and  construction of the LVDB and  some ideas for future expansion.
The LVDB will be useful for comparing new observational results or theoretical models  to the dwarf galaxy and star cluster populations in the  Local Volume.
I have presented several uses cases and examples along with public \texttt{jupyter} notebooks. 
The LVDB is publicly available as a GitHub repository\footnote{\url{https://github.com/apace7/local_volume_database}} and  community contributions are encouraged.
There are many upcoming surveys, such as the LSST at the Vera C. Rubin Observatory, the Euclid mission, and the Nancy Grace Roman Space Telescope, that will significantly expand our knowledge of the Local Volume and push the frontier  of  dwarf galaxy and star cluster research.
The upcoming era of near-field cosmology is bright and the LVDB is a tool built to enable research for this exciting era.

\section*{Acknowledgements}

I thank Jordan Bruce, Rachel Buttry, Jeff Carlin, William Cerny, Denija~Crnojevi\'c, Alex Drlica-Wagner, Chris Garling, Marla Geha, Alex Ji, Nitya Kallivayalil, Sergey Koposov, Ting S. Li, John Maner, David Nidever, Liam Plybon, Dave Sand, Josh Simon, Louie Strigari, Chin-Yi Tan, Erik Tollerud, Kathy Vivas,  Matt Walker, and Brian Yanny for helpful conversions about the LVDB, comments on the draft, and testing initial versions of the LVDB.
I thank the referee for their comments and feedback.

ABP acknowledgements  partial support by NSF grant AST-2206046.

This research has made use of NASA's Astrophysics Data System Bibliographic Services and the arXiv preprint server.
I acknowledge the usage of the HyperLeda database\footnote{\url{http://leda.univ-lyon1.fr/}} \citep{HyperLEDA_Makarov2014A&A...570A..13M}.

\section*{Software}

\texttt{astropy} \citep{Astropy2013A&A...558A..33A, Astropy2018AJ....156..123A},
\texttt{matplotlib} \citep{matplotlib}, 
\texttt{NumPy} \citep{numpy},
\texttt{iPython} \citep{ipython},
\texttt{SciPy} \citep{2020SciPy-NMeth}
\texttt{corner.py} \citep{corner}, 
\texttt{emcee} \citep{ForemanMackey2013PASP..125..306F},
\texttt{gala} \citep{gala},
\texttt{galpy}\footnote{ http://github.com/jobovy/galpy} \citep{Bovy2015ApJS..216...29B},

\bibliographystyle{aasjournal}
\bibliography{main_bib_file}

\appendix

\section{Detailed Description of the Input \texttt{YAML} Files}
\label{appendix:yaml}

The LVDB is structured a collection of\texttt{YAML} files with the properties of  each  individual  system collected in a single \texttt{YAML} file.  The \texttt{YAML} files are combined into user friendly catalogs. 
The LVDB uses  \texttt{YAML} files for several reasons: \texttt{YAML} files  are a  human readable file format, new systems  can easily be added to the database,  properties can be changed by editing a single file, and missing values do not need any input or null values. I note that not all of the \texttt{YAML} keys are included in the primary catalogs. 
\texttt{YAML} files are organized with keys and collections. Each \texttt{YAML} key  has an entry (numerical value, list, or string) and the \texttt{YAML} collections group \texttt{YAML} keys. 
The LVDB \texttt{YAML} file organizes various measurements as \texttt{YAML} collections (e.g., structure, distance, velocity). Each \texttt{YAML} collection generally has a single reference.
There are four  \texttt{YAML} keys  required for each system.
The detailed description of each \texttt{YAML} key are as follows:
\begin{itemize}
    \item \texttt{key} (required)---the unique internal identifier a  system in the LVDB, this also corresponds to the name of the \texttt{YAML} file. All LVDB \texttt{key}'s are lower case and most distant systems are grouped (for example, Cen~A satellites start with cen\_a\_ and the LMC globular clusters start with lmc\_gc\_).
    \item \texttt{table} (required)---the table to place the system in the combined catalogs. 
    \item \texttt{location} (required)---Collection with \texttt{YAML} keys: \texttt{ra} and \texttt{dec}. Location of system [degrees, ICRS frame, J2000.0].
    \item \texttt{name\_discovery}---Collection on discovery and classification information.
    \begin{itemize}
        \item \texttt{name}: Name of the system. 
        \item \texttt{other\_name}: List of additional names of the system. 
        \item \texttt{ref\_discovery}: List of discovery reference(s).  Multiple references are allowed to account for simultaneous discoveries, rediscovery, confirmation, and/or classification changes. 
        \item \texttt{discovery\_year}: The year of discovery.
        \item \texttt{host}: Host galaxy of the system.  
        \item \texttt{confirmed\_dwarf}: Dwarf galaxy classification column. 1 corresponds to dwarf galaxy classification. See Section~\ref{sec:classification} for details on the classification.
        \item \texttt{confirmed\_star\_cluster}: Star cluster classification column.  1 corresponds to a star cluster classification. See Section~\ref{sec:classification} for details on the classification.
        \item \texttt{confirmed\_real}: Denotes whether the system should be considered a candidate (=0) or whether the system has been confirmed (=1). Confirmation may occur with deeper photometric imaging, astrometry, and/or spectroscopy.
        For distant dwarf galaxies ($d\gtrsim 3~{\rm Mpc}$) this generally requires space based  imaging. 
        \item \texttt{false\_positive}: Denotes whether system is a false positive (=1; chance alignment or artifact) or background galaxy (=2). 
        \item \texttt{ref\_false\_positive}: Reference(s) that confirm the  system is a false positive or background galaxy.
        \item \texttt{type}: Type of galaxy or star cluster (e.g.,  dSph, dIrr, dE, GC, SC, NSC etc).
        \item \texttt{abbreviation}: Short 3-5 character abbreviation for the system.
    \end{itemize}
    \item \texttt{structure}---Collection of structural parameters. 
    \begin{itemize}
        \item \texttt{rhalf}: Projected (2D) half-light radius along the major axis of the system in angular units. The  units are specified in the \texttt{spatial\_unit} key. If the \texttt{spatial\_unit} is not included, the default units are arcmin.
        \item \texttt{spatial\_unit}: Input spatial units  of the \texttt{rhalf} column [options = arcmin or arcsec]. 
        \item \texttt{ellipticity}: Ellipticity of the system, defined as,  $\epsilon = 1 - b/a$, where b and a are the minor and major axis, respectively.
        \item \texttt{position\_angle}: Position angle of the system, defined as the angle from North to East. 
        \item \texttt{ref\_structure}: Reference.
    \end{itemize}
    \item \texttt{distance}---Distance collection. 
    \begin{itemize}
        \item \texttt{distance\_modulus}: Distance modulus.
        \item \texttt{distance\_fixed\_host}: When this option is set (=true), the distance to the system is fixed to its host.
        \item \texttt{ref\_distance}: Reference.
    \end{itemize}
    \item \texttt{m\_v}---Apparent magnitude collection.
    \begin{itemize}
        \item \texttt{apparent\_magnitude\_v}: Apparent magnitude in the V-band corrected for extinction.  The extinction correction is adopted from the literature source.
        \item \texttt{mean\_ebv}: Mean E(B-V) reddening.
        \item \texttt{ref\_m\_v}: Reference.
    \end{itemize}
    \item \texttt{velocity}---Stellar kinematic collection.
    \begin{itemize}
        \item \texttt{vlos\_systemic}: Systemic heliocentric (line-of-sight) velocity of the system [${\rm km~s^{-1}}$]. 
        \item \texttt{vlos\_sigma}: Global line-of-sight velocity dispersion [${\rm km~s^{-1}}$]. 
        \item \texttt{vlos\_sigma\_central}: Central velocity dispersion  [${\rm km~s^{-1}}$]. Generally used for globular clusters.
        \item \texttt{vlos\_rotation}: Stellar rotation [${\rm km~s^{-1}}$].
        \item \texttt{ref\_vlos}: Reference.
    \end{itemize}
    \item \texttt{proper\_motion}---Systemic proper motion collection.
    \begin{itemize}
        \item \texttt{pmra}: Systemic proper motion in right ascension [${\rm mas~yr^{-1}}$]. Includes $\cos{\delta}$ term.
        \item \texttt{pmdec}: Systemic proper motion in declination  [${\rm mas~yr^{-1}}$].
        \item \texttt{ref\_proper\_motion}: Reference.
    \end{itemize}
    \item \texttt{spectroscopic\_metallicity}---Stellar spectroscopic metallicity collection.
    \begin{itemize}
        \item \texttt{metallicity\_spectroscopic}: Mean spectroscopic metallicity ([Fe/H])
        \item \texttt{metallicity\_spectroscopic\_sigma}: Spectroscopic metallicity dispersion.
        \item \texttt{ref\_metallicity\_spectroscopic}: Reference.
    \end{itemize}
    \item \texttt{photometric\_metallicity}---Photometric metallicity collection.
    \begin{itemize}
        \item \texttt{metallicity\_photometric}: Metallicity measured from photometric studies (e.g.,  metallicity sensitive photometry).
        \item \texttt{metallicity\_photometric\_sigma}: Metallicity dispersion from photometric based measurements.
        \item \texttt{ref\_metallicity\_photometric}: Reference.
    \end{itemize} 
    \item \texttt{metallicity\_isochrone}---Photometric metallicity collection.
    \begin{itemize}
        \item \texttt{metallicity\_isochrone}: Metallicity measured from  isochrone or color-magnitude diagram. 
        \item \texttt{metallicity\_isochrone\_sigma}: Metallicity dispersion from isochrone or color-magnitude diagram based measurements
        \item \texttt{ref\_metallicity\_isochrone}: Reference.
    \end{itemize} 
    \item \texttt{structure\_king}---King stellar profile collection. \citep{King1962AJ.....67..471K}.
    \begin{itemize}
        \item \texttt{rcore}: King core radius.
        \item \texttt{rking}: King limiting radius. Commonly referred to as the tidal radius.
        \item \texttt{spatial\_unit}: Spatial units [arcmin or arcsec] of \texttt{rcore} and \texttt{rking}.
        \item \texttt{ellipticity}: Ellipticity from King profile fit.
        \item \texttt{position\_angle}: Position angle from King profile fit.
        \item \texttt{ref\_king}---Reference.
    \end{itemize}
    \item \texttt{structure\_sersic}---Sersic spatial profile collection \citep{Sersic1968adga.book.....S}.  The 2D Sersic profile used in the LVDB is defined  following \citet{Graham2005PASA...22..118G}: $\Sigma(R) = \exp\left[- b_n \left((R/R_s)^{1/n} -1\right) \right]$, where $R_s$ and $n$ correspond to \texttt{rad\_sersic} and \texttt{n\_sersic}, respectively and $b_n=1.9992~ n-0.3271$. 
    \begin{itemize}
        \item \texttt{n\_sersic}: Sersic power-law index. 
        \item \texttt{rad\_sersic}: Sersic scale radius. 
        \item \texttt{spatial\_unit}: Spatial units [arcmin or arcsec] of \texttt{rad\_sersic}.
        \item \texttt{ellipticity}: Ellipticity from Sersic profile fit.
        \item \texttt{position\_angle}: Position angle from Sersic profile fit.
        \item \texttt{ref\_sersic}: Reference.
    \end{itemize}
    \item \texttt{structure\_eff}---EFF spatial profile collection \citep{Elson1987ApJ...323...54E}.
    \begin{itemize}
        \item \texttt{gamma\_eff}: EFF power-law index. 
        \item \texttt{rad\_eff}: EFF scale radius. 
        \item \texttt{spatial\_unit}: Spatial units [arcmin or arcsec] of \texttt{rad\_eff}.
        \item \texttt{ellipticity}: Ellipticity from EFF profile fit.
        \item \texttt{position\_angle}: Position angle from EFF profile fit.
        \item \texttt{ref\_eff}: Reference.
    \end{itemize}
    \item \texttt{structure\_plummer}---Plummer spatial profile collection \citep{Plummer1911MNRAS..71..460P}.
    \begin{itemize}
        \item \texttt{rplummer}: Plummer profile scale radius. 
        \item \texttt{spatial\_unit}: Spatial units [arcmin or arcsec] of \texttt{rplummer}.
        \item \texttt{ellipticity}: Ellipticity from Plummer profile fit.
        \item \texttt{position\_angle}: Position angle from Plummer profile fit.
        \item \texttt{ref\_plummer}: Reference.
    \end{itemize}
    \item \texttt{structure\_exponential}---Exponential spatial profile collection.
    \begin{itemize}
        \item \texttt{rexponential}: Exponential scale radius. 
        \item \texttt{spatial\_unit}: Spatial units [arcmin or arcsec] of \texttt{rexponential}.
        \item \texttt{ellipticity}: Ellipticity from Exponential profile fit.
        \item \texttt{position\_angle}: Position angle from Exponential profile fit.
        \item \texttt{ref\_eff}: Reference.
    \end{itemize}
    \item \texttt{age}---Age collection. 
    \begin{itemize}
        \item \texttt{age}: Mean age of system [Gyr] for single stellar population systems (i.e., star clusters)
        \item \texttt{ref\_age}: Reference.
    \end{itemize}
    \item \texttt{flux\_HI}---Gas (HI)  collection.
    \begin{itemize}
        \item \texttt{flux\_HI}---HI flux of system [${\rm Jy~km~s^{-1}}$].
     \item \texttt{vlos\_systemic\_HI}---Systemic velocity of HI gas in system [${\rm km~s^{-1}}$].
     \item \texttt{sigma\_HI}---velocity dispersion of HI gas [${\rm km~s^{-1}}$].
     \item \texttt{vrot\_HI}---rotation velocity of HI gas [${\rm km~s^{-1}}$].
    \item \texttt{ref\_flux\_HI}---Reference.
    \end{itemize}
    \item \texttt{star\_formation\_history}: Star formation history derived from resolved star photometry and deep color-magnitude diagrams collection. Galaxy focused.  
    \begin{itemize}
        \item \texttt{tau\_50}: time for 50 per cent of the total stellar mass to form [Gyr]. Median age of system.
        \item \texttt{tau\_80}: time for 80 per cent of the total stellar mass to form [Gyr]. Quenching Time. 
        \item \texttt{tau\_90}: time for 90 per cent of the total stellar mass to form [Gyr]. Quenching Time. 
        \item \texttt{ref\_star\_formation\_history}: Reference.
    \end{itemize}
\item \texttt{notes}---Miscellaneous notes about the system.  This may include details about measurements in the LVDB \texttt{YAML} file or details about the classification of the system. The notes vary significantly from system to system.
\end{itemize}

For systems without a particular measurement, the corresponding \texttt{YAML} key is not included and the parameters will be empty in the combined catalog. 
For all measured parameters there can be error keys. The errors keys are column name + (\texttt{\_em}, \texttt{\_ep}, \texttt{\_ul}, \texttt{\_ll}). The error names correspond to: \texttt{\_em}---16\% confidence/credible interval (error minus; i.e., minus one sigma); \texttt{\_ep}---84\% confidence/credible interval (error plus; i.e., plus one sigma); \texttt{\_ul}---95.5\% confidence/credible interval (upper limit, i.e. two sigma upper limit), generally used as an upper limit; \texttt{\_ll}---4.5\% confidence/credible interval (lower limit, i.e. two sigma lower limit), generally used as an lower limit. 
For some entries (generally, the velocity or metallicity dispersion), I have reanalyzed data sets to quote consistent upper limits. However, even with this reanalysis some limits are quoted at different values\footnote{The most common measurement with different limits are HI flux limits. These vary from 2-5 $\sigma$ in the literature.}.  When this occurs, this detail are included in the \texttt{notes} key.

The \texttt{ref} keys follow a consistent schema of author last name + ADS bibcode.  The author's last name is stripped of special characters and spaces but capitalization is kept. While the ADS bibcode is a unique identifier for a reference, the addition of the author's last name significantly improves the human interpretability of the LVDB reference columns.
To aid in enabling references to the internal contents of the database, there is an up-to-date BibTeX file containing all entries in the LVDB with the LVDB \texttt{ref} schema\footnote{ \href{https://github.com/apace7/local_volume_database/blob/main/table/lvdb.bib}{Link to BibTeX file}} and I have made an ADS library that  includes the LVDB references\footnote{\href{https://ui.adsabs.harvard.edu/public-libraries/fVKkEJbdRyCmscCOwzsz6w}{Link to ADS library}}.
As the ADS bibcode has a fixed length of 19 characters, the ADS bibcode can be derived from the last 19 characters of the LVDB \texttt{ref} columns.

For many systems,  there may be a number of different studies of the same measurement or property (i.e. distance, half-light radius etc) and the decision of which measurement to include is subjective. The sample size, depth, field-of-view, precision, and methodology are considered when deciding on which measurement to include. 
I note that many \texttt{YAML} keys were added after the creation of the  LVDB  and many newer entries are incomplete  (this includes \texttt{mean\_ebv}, \texttt{type}, HI kinematics, \texttt{star\_formation\_history}).

There are some fundamental dwarf galaxy and star cluster properties that I have opted not to include.  These   quantities  are generally model dependent.  One  example are the orbital properties (e.g., pericenter and apocenter) of dwarf galaxies and star clusters. The orbital properties are highly dependent on the MW potential assumed, whether the influence of the LMC was included in the orbit calculations, the  input parameters (which may differ from the values here), and the choice of solar motion. 
While I include the dynamical mass within the half-light radius as a value-added quantity, I have not included any  mass or density measurements from detailed dynamical modeling.

\subsection{Value-Added Columns}
\label{appendix:value}

The combined catalogs have additional value-added columns and these include:

\begin{itemize}
\item \texttt{M\_V}---Absolute magnitude in the V-band ($M_V$).  Computed from the apparent magnitude and distance modulus.  Note that the distance errors are commonly not included when computing the the absolute magnitude and I follow that convention here.
\item \texttt{mass\_stellar}---Stellar mass computed from $M_V$ assuming a mass-to-light ratio of 2 [$\log_{10}{M_{\odot}}$]. 
\item \texttt{distance}---Heliocentric distance to the system, computed from the distance modulus [kpc].
\item \texttt{ll}, \texttt{bb}---Location of the system in Galactic coordinates [degree, degree].
\item \texttt{sg\_xx}, \texttt{sg\_yy}, \texttt{sg\_zz}---Cartesian coordinates of the system in Supergalactic coordinates [kpc, kpc, kpc].
\item \texttt{distance\_gc}, \texttt{distance\_m31}, \texttt{distance\_lg}, \texttt{distance\_host}---Distance of the system relative to the Galactic center, M31, the Local Group center, and the system's host galaxy, respectively [kpc, kpc, kpc, kpc].
\item \texttt{rhalf\_physical}---Major-axis of half-light radius in physical units [parsec]. This column includes errors from a Monte Carlo sampling of the half-light radius and distance errors.
\item \texttt{rhalf\_sph\_physical}---azimuthally-averaged half-light radius (also referred to as the geometric mean) in physical units [parsec]. This column includes errors from a Monte Carlo sampling of the half-light radius, ellipticity, and distance errors.
\item \texttt{surface\_brightness\_rhalf}---Average surface brightness enclosed within the azimuthally-averaged half-light radius [${\rm mag~arcsec^{-2}}$].
\item \texttt{mass\_HI}---HI (gas) mass. The HI mass is computed from $M_{HI}=2.356\times 10^5  (d/{\rm Mpc})^2~S$, where $S$ is the HI flux and $d$ is the distance in Mpc [$\log_{10}{M_{\odot}}$].  
\item \texttt{metallicity}---Combined metallicity column. The order of preference is: spectroscopic, photomteric, and isochrone. 
\item \texttt{metallicity\_type}---Metallicity type (spectroscopic, photometric, isochrone) of \texttt{metallicity} column.
\item \texttt{velocity\_gsr}---Velocity in the Galactic standard of rest [${\rm km~s^{-1}}$].
\item \texttt{velocity\_lg}---Velocity relative to the Local Group [${\rm km~s^{-1}}$]
\item \texttt{mass\_dynamical\_wolf}---Dynamical mass using the \citet{Wolf2010MNRAS.406.1220W} mass estimator for a pressure-supported system ($M(r=r_{1/2})=930~\sigma^2~R_{1/2}$).  This column includes errors from a Monte Carlo sampling of the velocity dispersion, half-light radius, ellipticity, and distance errors [$\log_{10}{M_{\odot}}$].
\end{itemize}

\section{Secondary Compilations}

The primary catalogs in the LVDB   only include a single entry per property. There are two additional catalogs that are included as compilations of literature measurements. 

\subsection{Systemic Proper Motion of Dwarf Galaxies}

I include a compilation of systemic proper motion measurements of dwarf galaxies and some HFCSS \footnote{\href{https://github.com/apace7/local_volume_database/blob/main/data/pm_overview.csv}{Link to the proper motion compilation.}}. 
This catalog was created for literature comparison in the \citet{Pace2022ApJ...940..136P} analysis\footnote{The comparison figures are only included in the arXiv/perprint version of \citet{Pace2022ApJ...940..136P}.} and subsequently updated. 
This includes ground, space based (e.g., {\it HST}), and {\it Gaia} measurements. 
Each entry in the catalog contains: LVDB key, LVDB reference, ADS bibcode, systemic proper motion measurement ($\mu_{\alpha \star}$, $\mu_\delta$, including the correlation and asymmetric errors), text citation (e.g., Pace et al. 2022 or Simon 2020), method (Gaia, HST, etc), and comments.  
There is an example notebook\footnote{\href{https://github.com/apace7/local_volume_database/blob/main/example_notebooks/example_pm_overview.ipynb}{Link to example proper motion notebook.}} in the GitHub repository that describes the use of this catalog and creates comparison figures of systemic proper motion measurements for a system. These plots are similar to comparison plots in \citet{Pace2022ApJ...940..136P} and can be used to compare new measurements to the literature. 
This compilation is  complete for MW dwarf galaxies and  only contains published measurements.

\subsection{Astrophysical J-factors for the Indirect Detection of Dark Matter}

I include a compilation of astrophysical J-factors of MW dwarf galaxies for studies of the indirect detection of dark matter\footnote{\href{https://github.com/apace7/local_volume_database/blob/main/data/j_factor.csv}{Link to J-factor compilation.}}.
The  J-factor is the astrophysical component of the dark matter annihilation flux.
There are a number of J-factors analyses focusing on the MW dwarf galaxies \citep[e.g.,][]{GeringerSameth2015ApJ...801...74G, Bonnivard2015MNRAS.453..849B, Pace2019MNRAS.482.3480P}. 
Similar to the proper motion compilation, this is a compilation of measurements and includes multiple measurements per galaxy.
Each entry in the catalog contains: LVDB key, LVDB reference, ADS bibcode, text citation, selection (i.e., methodology grouping), angle [degree], J-factor  measurement [($\log_{10}{J}$] (this includes median, one and two sigma asymmetric errors and an upper limit column), use (another selection type column), and comments.
I include this compilation primarily to compare new J-factor measurements to literature results and in many analyses, utilizing a consistent J-factor analysis is ideal.
Unlike the proper motion compilation, the J-factor  compilation is quite incomplete of literature analyses.
There is an example notebook\footnote{\href{https://github.com/apace7/local_volume_database/blob/main/example_notebooks/example_j_factor.ipynb}{Link to example J-factor notebook.}} in the GitHub repository that describes the use of this catalog and creates some summary figures.

\section{Summary Tables}

In addition to data catalogs, there is a pdf document with summary tables of the LVDB content. These tables are created via python and LaTeX scripts from the LVDB catalogs and \texttt{YAML} files and the summary file is included as part of the GitHub release\footnote{Named \texttt{lvdb\_table.pdf} on the GitHub release pages \href{https://github.com/apace7/local_volume_database/releases}{(Link)}.}.
Each primary catalog has four tables and the content of table each is  described below.

The first table is focused on the discovery and classification of each system. The columns are as follows: (1) System name. (2) Other names for system. (3) RA (hms). (4) Dec (dms). (5) Host Galaxy. (6) Discovery reference(s). (7) Denotes whether system is confirmed  or a candidate.  (8) Type/classification of the system. This includes both dwarf galaxy (dSph, dIrr, etc) and star cluster types (star cluster (SC), nuclear star cluster (NSC), globular cluster (GC)). 
This table is sorted by discovery year and then LVDB key. For the dwarf galaxy hosted globular clusters and Local Volume dwarf galaxy tables, the systems are first sorted by host LVDB key.

The second table is focused on the structural parameters, distance, and luminosity. The columns are as follows: (1) System name. (2) RA (deg). (3) Dec (deg). (4)  Half-light radius along the major axis. (5) Ellipticity. (6) Position angle. (7) Azimuthally-averaged half-light radius in physical units. (8) Distance modulus. (9) Distance. (10) Extinction corrected apparent V-band magnitude. (11) Extinction corrected  absolute V-band magnitude. (12) References.
This table and the following tables are sorted by the LVDB key. For the dwarf galaxy hosted globular cluster and Local Volume  dwarf galaxy tables, this table and the following tables are first sorted by host LVDB key then system LVDB key.

The third table is focused on the systemic motion, (stellar) kinematics, and (spectroscopic) metallicity. The columns are as follows: (1) System name. (2) Galactic longitude (deg). (3) Galactic latitude (deg). (4) Systemic (line-of-sight) velocity. (5) Global velocity dispersion. (6) Spectroscopic  stellar mean  metallicity. (7) Spectroscopic metallicity dispersion.  (8) Systemic proper motion in the direction of right ascension. (9) Systemic proper motion in the direction of  declination. (10) References. For globular cluster tables, an additional age column is added after the spectroscopic metallicity dispersion column.

The fourth table is focused on the stellar mass, dynamical mass, and HI gas mass. The columns are as follows: (1) System name. (2) Stellar mass assuming a  mass-to-light ratio of 2. (3) Dynamical mass within the 3D half-light radius from stellar kinematics.  (4) Dynamical mass-to-light ratio within the half-light radius from stellar kinematics. (5) HI gas mass. (6) Ratio of HI-to-stellar mass. (7) References. This table is not included for the globular clusters catalogs.

Lastly, there are two tables for low significance candidates, false positive, and background galaxy  systems.  The first table is similar to the previous discovery tables with the candidate column is replaced by a description of why the system is included in this catalog. This includes: candidate (Cand.), known false positive (FP), or background galaxy (BG). The type column is replaced with references to the false positive classification.
The second table is similar to the structural parameter table, however, I omit errors and add the Galactic longitude and latitude.

\end{document}